\documentclass[aps, prb, 10pt, twocolumn, superscriptaddress, notitlepage]{revtex4-2}
\usepackage{natbib}
\usepackage{graphicx}
\usepackage{dcolumn} 
\usepackage{bm}
\usepackage{mathtools}
\usepackage{amssymb}
\usepackage{enumitem}
\usepackage[svgnames,dvipsnames]{xcolor}
\usepackage[normalem]{ulem}
\usepackage[mathlines]{lineno}
\usepackage[titletoc, title]{appendix}
\usepackage{bbold}
\usepackage{comment}
\usepackage{mathrsfs}
\usepackage{hyperref}
    \hypersetup{unicode=true,
        pdftoolbar=true,
        pdfmenubar=true,
        pdffitwindow=false,
        pdfstartview={FitH},
        pdfnewwindow=true,
        colorlinks=true,
        linkcolor=blue,
        citecolor=blue,
        filecolor=blue,
        urlcolor=blue}

\newcommand{\beq}{\begin{equation}}
\newcommand{\eeq}{\end{equation}}
\newcommand{\beqa}{\begin{eqnarray}}
\newcommand{\eeqa}{\end{eqnarray}}

\begin{document}

\title{Quantum effective action for dissipative semiclassical dynamics}

\author{Cesare Vianello}
\email{cesare.vianello@phd.unipd.it}
\affiliation{\mbox{Dipartimento di Fisica e Astronomia ``Galileo Galilei'', Università di Padova, via Marzolo 8, 35131 Padua, Italy}}
\affiliation{Istituto Nazionale di Fisica Nucleare, Sezione di Padova, via Marzolo 8, 35131 Padua, Italy}
\author{Andrea Bardin}
\affiliation{\mbox{Dipartimento di Fisica e Astronomia ``Galileo Galilei'', Università di Padova, via Marzolo 8, 35131 Padua, Italy}}
\affiliation{Istituto Nazionale di Fisica Nucleare, Sezione di Padova, via Marzolo 8, 35131 Padua, Italy}
\author{Luca Salasnich}
\affiliation{\mbox{Dipartimento di Fisica e Astronomia ``Galileo Galilei'', Università di Padova, via Marzolo 8, 35131 Padua, Italy}}
\affiliation{Istituto Nazionale di Fisica Nucleare, Sezione di Padova, via Marzolo 8, 35131 Padua, Italy}
\affiliation{Padua QTech Center, Università di Padova, via Gradenigo 6/A, 35131 Padua, Italy}

\begin{abstract}
    Using the quantum effective action in the Schwinger-Keldysh formalism, we derive quantum-thermal corrections to the semiclassical dynamics of a dissipative system described by a macroscopic degree of freedom. We discuss the connection with the Ehrenfest theorem and, within the local potential approximation, obtain the corrections in closed form for arbitrary temperature and damping, showing that as the temperature increases, they cross over from purely quantum to purely thermal. In the low-temperature and weak-damping regime, corrections are set by the zero-point energy of fluctuations evaluated at the classical underdamped frequency, closely paralleling the conservative case, which allows us to go to higher order in the derivative expansion. We apply these general results to the resistively and capacitively shunted superconducting Josephson junction and to an elongated bosonic junction, where corrections can reach the percent level under realistic conditions.
\end{abstract}

\maketitle

\section{Introduction}

Superfluid states of matter are characterized by macroscopic quantum coherence. Some of their low-energy properties may thus be described at the mean-field level in terms of a complex-valued order parameter, a classical field that encodes the collective behavior of the underlying microscopic degrees of freedom and is governed by nonlinear field equations, such as the Gross-Pitaevskii and Ginzburg-Landau equations \cite{landau_book, svistunov_book}. In particular, the Josephson effect in tunneling junctions \cite{josephson1962, anderson1963, barone_book, smerzi1997, milburn1997, cataliotti2001, albiez2005, gati2007, levy2007, sakmann2009, spagnolli2017, singh2018, bernhart2025, gammal2026} can be understood as the dynamics of the relative phase between the order parameters of two coupled superfluids, which constitutes a single macroscopic degree of freedom. 

The classical conservative dynamics for the phase, although adequate in many scenarios, falls short of capturing two key aspects. First, in several experimental realizations the actual dynamics is dissipative. In superconducting junction circuits, dissipation arises from the presence of external resistors \cite{barone_book, kagan_book, larkin_book, furutani2021}, while in bosonic junctions it originates from the intrinsic coupling between the Josephson mode and the phonon bath of the condensate \cite{marino1999, meier2001, barankov2003, pigneur2018, pigneur2018b, polo2018, stefanatos2019, saha2020, xhani2020, binanti2021, mennemann2021, xhani2022, saha2023, saha2026}. Second, the phase has a quantum nature that manifests in beyond-mean-field effects. In superconducting junctions, this was established in pioneering experiments by Clarke \emph{et al.}~\cite{koch1980, koch1981, koch1982, martinis1985, devoret1985, devoret1988}, which observed zero-point fluctuations, macroscopic quantum tunneling, and energy-level quantization of the phase. In bosonic junctions, where the phase is directly related to the condensate fraction, its quantum nature is reflected in the interaction-induced quantum depletion \cite{vianello2025a, vianello2026}.

In recent works \cite{furutani2022, vianello2025b, salasnich2025}, we suggested that the quantum effective action, originally introduced to study quantum effects in relation to spontaneous symmetry breaking \cite{goldstone1962, jona1964, coleman1973, coleman1974}, provides a transparent framework to incorporate quantum corrections to the mean-field dynamics of the order parameter. Moreover, it offers a systematic interpretation of effective approaches such as the modified Gross-Pitaevskii equation, which is often introduced heuristically in the spirit of density functional theory and has found numerous applications in recent years (see, e.g., Refs.~\cite{petrov2015, bisset2016, salasnich2018, aybar2019, guebli2021, bardin2024}). At first order in $\hbar$, the quantum-corrected semiclassical dynamics may be derived analytically; applied to Josephson junctions in the absence of dissipation, it was shown to systematically improve upon the mean field when compared with exact quantum results \cite{vianello2025a, vianello2025b}.

In this work, we incorporate dissipation into the quantum effective action, thereby characterizing its interplay with quantum fluctuations of the system and addressing jointly both of the key aspects outlined above. To this end, we adopt the Schwinger-Keldysh formalism \cite{schwinger1961, keldysh1965}, which provides a consistent action principle for describing dissipative dynamics. In this framework, dissipation and stochastic noise are intrinsically linked through the fluctuation-dissipation theorem \cite{callen1951, kubo1966}, reflecting the physical fact that dissipation originates from the coupling to environmental degrees of freedom (assumed to be in equilibrium) which are themselves subject to quantum and thermal fluctuations; these fluctuations, in turn, induce a stochastic force on the system. Our aim is then to derive quantum corrections to the deterministic component of the dissipative dynamics, which arise from fluctuations of the system's collective degree of freedom coupled to the environment. We will show that such corrections generally depend on $\hbar$, the damping coefficient $\gamma$, and the temperature $T$, and that, as the temperature increases, they cross over from purely quantum to purely thermal character. In certain limiting regimes, they admit a transparent physical interpretation in terms of the energy of quantum-thermal fluctuations evaluated at a characteristic frequency.

The paper is organized as follows. In Sec.~\ref{sec:theory}, we derive the one-loop Keldysh quantum effective action for a macroscopic degree of freedom, emphasizing the connection of this approach to the Ehrenfest theorem, the distinction between quantum and classical corrections, and the role of fluctuations around the corrected average. We discuss the local potential approximation and, in the low-temperature and weak-damping regime, the second-order derivative expansion. Moreover, we extend the results to the case in which the conservative part of the action is formulated in phase space. In Sec.~\ref{sec:applications}, we apply the general results to the resistively and capacitively shunted junction (RCSJ) model for a superconducting circuit and to an elongated bosonic junction, which provide concrete realizations of the two types of action considered in the previous section. Finally, in Sec.~\ref{sec:conclusions} we present our conclusions and perspectives on possible developments, commenting on the connection of our results to vacuum-instability phenomena, the Schwinger effect and the Casimir effect. Appendixes \ref{app:LPA}, \ref{app:B}, and \ref{app:C} provide detailed calculations of the corrections in the local potential approximation, the derivation of the Keldysh action for the elongated bosonic junction, and an alternative derivation of the quantum corrections in the low-temperature and weak-damping regime via a generalized Hamilton principle.

\section{Dissipative quantum action}\label{sec:theory}

\subsection{Keldysh action}

We consider our macroscopic degree of freedom as the coordinate $x(t)$ of a particle of mass $m>0$ moving in a potential $U(x)$. Its conservative dynamics follows from Hamilton's least action principle for
\begin{equation}\label{eq:s0}
    S_0[x]=\int dt \left[\frac{m}{2}\dot x^2 - U(x)\right].
\end{equation}
How to describe dissipation? A generic dissipative dynamics does not admit an ordinary variational formulation in terms of a single dynamical variable and a standard action $S[x]$ \cite{bauer1931}. Variational realizations typically require auxiliary variables \cite{bateman1931}, explicit time dependence \cite{caldirola1941, kanai1948}, or other nonstandard constructions, such as fractional derivatives \cite{riewe1, riewe2}. A consistent causal variational framework is provided by the Schwinger-Keldysh closed-time-path formalism \cite{schwinger1961, keldysh1965, kamenev_book, uguccioni2025}, in which the dynamics is formulated on doubled fields, namely forward and backward trajectories $x_+(t)$ and $x_-(t)$, or classical and response fields $x = (x_++x_-)/2$ and $x_q = x_+-x_-$. The Keldysh action is written as
\begin{equation}\label{eq:Kac}
    S[x, x_q] = S_\text{cons}[x, x_q] + S_\text{diss}[x, x_q],
\end{equation}
where
\begin{align}
    S_\text{cons} &= S_0[x_+] - S_0[x_-] \nonumber\\
    &= \int dt\left[-m\ddot x-U'(x)\right]x_q+\mathscr O(x_q^3)
\end{align}
encodes the conservative part of the dynamics, while
\begin{align}\label{eq:acK_diss}
    S_\text{diss} = \int dt\,dt'&\biggl[x_q(t)\alpha_R(t-t')x(t')\nonumber\\
    &+ \frac{i}{2}x_q(t)\alpha_K(t-t')x_q(t')\biggr]
\end{align}
accounts for dissipation. Here, $\alpha_R(t-t')$ denotes the retarded response kernel, while $\alpha_K(t-t')$ is the Keldysh correlation kernel. Both emerge from the microscopic action describing the particle coupled to a quantum-thermal bath after integrating out the bath degrees of freedom, see, e.g., Appendix \ref{app:B}. If the bath is in thermal equilibrium, the two kernels are related by the fluctuation-dissipation theorem \cite{callen1951, kubo1966}
\begin{equation}\label{eq:FT}
    \alpha_K(\omega)=\hbar \coth\left(\frac{\beta\hbar\omega}{2}\right)\text{Im}\,\alpha_R(\omega),
\end{equation}
where $\beta = (k_BT)^{-1}$ is the inverse temperature.

The quadratic term in $x_q$ can be decoupled via the Hubbard-Stratonovich (HS) transformation
\begin{equation}\label{eq:HS}
    e^{iS} = \int \mathcal D\xi\,e^{-\frac{1}{2}\int dt\, dt' \xi(t)\alpha_K^{-1}(t-t')\xi(t')}e^{i\mathcal S},
\end{equation}
where
\begin{equation}
    \mathcal S =\! \int\!dt\left[-m\ddot x \!-\! U'(x) \!+\!\! \int\!dt'\alpha_R(t-t')x(t') \!+\! \xi(t)\right]\! x_q
\end{equation}
is the Martin-Siggia-Rose (MSR) action \cite{martin1973, janssen1976, dedominicis1976, furutani2023}. The classical equation of motion (EoM) then follows from the saddle-point condition
\begin{equation}
    \frac{\delta\mathcal S}{\delta x_q}\biggr|_{x_q=0}=0,
\end{equation}
which yields the generalized Langevin equation
\begin{equation}\label{eq:Langevin}
    m\ddot x + U'(x) - \int dt'\alpha_R(t-t')x(t')=\xi(t).
\end{equation}
From Eq.~\eqref{eq:HS}, $\xi(t)$ is a Gaussian-distributed fluctuating force with $P[\xi] \propto \exp[-\frac{1}{2}\int dt\int dt'\xi(t)\alpha_K^{-1}(t-t')\xi(t')]$, implying $\langle \xi(t)\rangle=0$ and $\langle \xi(t)\xi(t')\rangle=\alpha_K(t-t')$. We thus identify the Keldysh correlator as the noise correlator.

From now on, we will take 
\begin{equation}\label{eq:ohm}
    \alpha_R(t-t')=-\gamma\,\partial_t\delta(t-t'),\qquad \gamma>0.
\end{equation}
This expression should be interpreted as a distributional shorthand for a kernel satisfying $\text{Im}\,\alpha_R(\omega)=\gamma\omega$, which reproduces linear Markovian damping originating from an Ohmic bath. In fact, since ${\partial_t\delta(t-t') = -\partial_{t'}\delta(t-t')}$, substituting Eq.~\eqref{eq:ohm} into Eq.~\eqref{eq:Langevin} and integrating by parts, we obtain
\begin{equation}\label{eq:eom_class}
m\ddot x + U'(x) + \gamma\dot x = \xi(t),
\end{equation}
with noise spectrum [Eq.~\eqref{eq:FT}]
\begin{equation}\label{eq:noise}
    \alpha_K(\omega) = \gamma\hbar\omega\coth\left(\frac{\beta\hbar\omega}{2}\right).
\end{equation}
This has two relevant limits: $\alpha_K(\omega) \simeq \gamma\hbar|\omega|$ in the low-temperature limit $\beta^{-1} \ll \hbar |\omega|$, describing purely quantum noise, and $\alpha_K \simeq 2\gamma k_B T$ in the high-temperature limit $\beta^{-1} \gg \hbar|\omega|$, describing purely thermal noise.

\subsection{Quantum effective action}

The main purpose of the paper is to compute quantum corrections to Eq.~\eqref{eq:eom_class}. To this end, we start from the corresponding Keldysh action [Eq.~\eqref{eq:Kac}]
\begin{align}\label{eq:keldysh_markov}
    S &= \int dt \left[-m\ddot x-\gamma\dot x-U'(x)\right]x_q\nonumber\\
    &+\frac{i}{2}\int dt\,dt'x_q(t)\alpha_K(t-t')x_q(t')\nonumber\\
    & -\frac{1}{24}\int dt\,U'''(x) x_q^3  + \mathscr O(x_q^5),
\end{align}
with $\alpha_K(t-t')$ the inverse Fourier transform of \eqref{eq:noise}. We expand the fields $x$, $x_q$ around background fields $\bar x$, $\bar x_q$ as $x=\bar x+\eta$ and $x_q=\bar x_q+\eta_q$. To second order in the fluctuations $\bm\eta = (\eta, \eta_q)$, the action reads 
\begin{align}
    S &\simeq S[\bar x, \bar x_q] +\frac{1}{2}\int dt\,dt'\bm\eta^T(t)\mathcal F(t-t')\,\bm\eta(t'),
\end{align}
where
\begin{gather}
    \mathcal F = \begin{pmatrix}
        -U'''(\bar x)\bar x_q + \mathscr O(\bar x_q^3) & \mathcal M^\dag +\mathscr O(\bar x_q^2) \\ \mathcal M +\mathscr O(\bar x_q^2) & i\alpha_K - \frac{1}{4}U'''(\bar x)\bar x_q+\mathscr O(\bar x_q^3)
    \end{pmatrix},\label{eq:F}\\
    \mathcal M = \delta(t-t')[-m\partial_t^2 - \gamma\partial_t - U''(\bar x)].\label{eq:M}
\end{gather}
The one-loop one-particle irreducible (1PI) quantum effective action $\Gamma[\bar x, \bar x_q]$ is then obtained by performing the Gaussian path integral over $\bm\eta$ \cite{burgess_book, kleinert_book, salasnich2025, furutani2022, vianello2025b, jackiw1974, iliopoulos1975, cametti2000}, which yields
\begin{equation}\label{eq:qea}
\Gamma[\bar x, \bar x_q] = S[\bar x, \bar x_q] + \frac{i}{2}\mathrm{Tr}\ln \mathcal{F}.
\end{equation}
Defining $\mathcal F_0 = \mathcal F|_{\bar x_q=0}$ and $\mathcal F_q = -U'''(\bar x)\bar x_q\,\text{diag}(1, \frac{1}{4})$, we expand the trace-log in series of $\bar x_q$ as $\text{Tr}\ln \mathcal F = \text{Tr}\ln(\mathcal F_0+\mathcal F_q)= \text{Tr}\ln\mathcal F_0+\text{Tr}(\mathcal F_0^{-1}\mathcal F_q) + \mathscr O(\bar x_q^2)$. The first term is independent of $\bar x_q$ (in fact, it is independent of both background fields; this is the normalization condition $\Gamma[\bar x, \bar x_q = 0]=0$, which expresses conservation of probability) and thus does not contribute to the EoM. The second term, which is linear in $\bar x_q$, will affect the deterministic part of the EoM. Note that since the 22 component of $\mathcal F_0^{-1}$ vanishes, the $\mathscr O(x_q^3)$ vertex of Eq.~\eqref{eq:keldysh_markov}---which at the order we retain contributes precisely to the 22 entry of $\mathcal F_q$---is annihilated in the trace and does not affect the result. Finally, the $\mathscr O(\bar x_q^2)$ term is generated by the system's own nonlinearity acting on its fluctuations. It is not a renormalization of the bath kernel $\alpha_K$, which is fixed exactly by the fluctuation-dissipation relation \eqref{eq:noise}, but rather an additional $\bar x$-dependent (multiplicative) noise kernel. Relative to the tree-level noise $\alpha_K$, it is of order $\mathscr O(\hbar)$, i.e.~the same order as the correction we retain in the drift; it therefore represents the leading quantum correction to the diffusive part of the dynamics and does not affect the deterministic evolution at the order considered here. Since our aim is the quantum-corrected deterministic dynamics, we shall drop it, together with the higher-order terms.

The quantum effective action is therefore $\Gamma = S + \Gamma_1$, with $\Gamma_1 = \frac{i}{2}\text{Tr}(\mathcal F_0^{-1} \mathcal F_q)$. Computing the trace, we obtain
\begin{equation}\label{eq:gamma1}
    \Gamma_1 =  -\frac{1}{2}\int dt\bigl[U'''(\bar x) \sigma(\bar x)\bigr]\bar x_q,
\end{equation}
where
\begin{equation}\label{eq:def_sigma}
    \sigma(\bar x) = \int\frac{d\omega}{2\pi}\frac{\alpha_K(\omega)}{|\mathcal M(\omega)|^2}.
\end{equation}
The 1PI saddle-point condition
\begin{equation}\label{eq:saddle}
    \frac{\delta\Gamma}{\delta\bar x_q}\biggr|_{\bar x_q=0}=0,
\end{equation}
which identifies the background field $\bar x$ as the average $\langle\hat x\rangle$, then yields the quantum-corrected EoM
\begin{equation}\label{eq:eom_corr}
    m\ddot{\bar x} + \gamma\dot{\bar x}+U'(\bar x)+\frac{1}{2}U'''(\bar x)\sigma(\bar x)= 0.
\end{equation}

Some comments are in order:

\emph{i. Connection to the Ehrenfest theorem}. As an approach to the semiclassical limit, the quantum effective action is closely related to the Ehrenfest theorem. 

Let us consider first the conservative case, for which the Heisenberg equation is $m\ddot{\hat x} +U'(\hat x)=0$. The Ehrenfest theorem then gives $m\ddot{\bar x}+\langle U'(\hat x)\rangle=0$, where $\bar x = \langle \hat x\rangle$ and the expectation value is understood in the vacuum associated with fluctuations around $\bar x$. Writing $\hat x = \bar x + \hat \eta$, with  $\langle \hat \eta \rangle=0$ by definition, and expanding the potential around $\bar x$, one obtains $\langle U'(\hat x)\rangle=U'(\bar x)+\frac{1}{2} U'''(\bar x)\langle \hat \eta^2\rangle + \mathscr O(\hat \eta^3)$. Comparing this with the conservative version ($\gamma = 0$) of Eq.~\eqref{eq:eom_corr}, for which $\sigma(\bar x) = \frac{\hbar}{2}[mU''(\bar x)]^{-1/2}$ \cite{furutani2022, vianello2025b, salasnich2025}, we realize that the two dynamics coincide provided that $\langle \hat \eta^2\rangle=\sigma(\bar x)$, meaning that the fluctuations $\hat\eta$ are identified with those of the ground state of a harmonic oscillator with frequency
\begin{equation}
\omega_0(\bar x)=\sqrt{\frac{U''(\bar x)}{m}}.
\end{equation}   
At the one-loop level this is exactly the case: the fluctuations over which one integrates to obtain the one-loop quantum effective action are Gaussian, with frequency $\omega_0(\bar x)$ \cite{furutani2022, vianello2025b, salasnich2025}.

In the dissipative case, we start from the quantum Langevin equation $m\ddot{\hat x} + \gamma \dot{\hat x} + U'(\hat x) = \hat\xi(t)$ \cite{metiu1984, ford1987, dittrich_book, hanggi_langevin, weiss_book, burmistrov_book}. The Ehrenfest theorem then yields $m\ddot{\bar x} + \gamma \dot{\bar x} + \langle U'(\hat x)\rangle = m\ddot{\bar x} + \gamma \dot{\bar x} + U'(\bar x)+\frac{1}{2} U'''(\bar x)\langle \hat \eta^2\rangle + \mathscr O(\hat \eta^3)=0$, where $\langle \,.\, \rangle$ now denotes the expectation value with respect to the reduced state of the system, obtained after tracing out the bath degrees of freedom. Substituting $\hat x = \bar x + \hat \eta$ and expanding to leading order in $\hat \eta$, we obtain $m\ddot{\hat \eta} + \gamma \dot{\hat\eta} + U''(\bar x)\hat \eta = \hat\xi(t)$, that is the equation for a driven-dissipative harmonic oscillator. The variance of $\hat \eta$ is then $\langle \hat \eta^2\rangle=\int\frac{d\omega}{2\pi}|\chi(\omega)|^2\alpha_K(\omega)$, where $\chi(\omega) = [-m\omega^2-i\gamma\omega+U''(\bar x)]^{-1}$ is the response function and $\alpha_K(\omega)$ is given by Eq.~\eqref{eq:noise}. Since $|\chi(\omega)|^2=\text{Im}\,\chi(\omega)/\gamma\omega$, this is consistent with the fluctuation-dissipation theorem for the system, $\langle \hat \eta^2\rangle=\int\frac{d\omega}{2\pi} \hbar\coth(\frac{\hbar\omega}{2k_BT})\text{Im}\,\chi(\omega)$. Finally, noting that $\chi(\omega) = -\mathcal M^{-1}(\omega)$, we immediately get $\langle\hat\eta^2\rangle = \sigma(\bar x)$, recovering exactly Eq.~\eqref{eq:eom_corr}. This establishes the formal correspondence between the one-loop quantum effective action and the leading-order correction derived from the Ehrenfest theorem. 

\emph{ii. Quantum vs classical}. Although we have been thinking of $\sigma(\bar x)$ as a quantum correction, namely, of $\bar x$ as the quantum average $\langle \hat x \rangle$, we emphasize that nothing in the derivation of this section is exclusively quantum. The `quantumness' of the bath is contained in Eq.~\eqref{eq:noise} for the noise kernel, and our results carry over unchanged in its classical limit, $\alpha_K \to 2\gamma k_B T$. Indeed, $\sigma(\bar x)=\langle\hat\eta^2\rangle$ is the total variance of the fluctuations around the background, to which zero-point motion and thermal occupation contribute on the same footing: only at low temperatures, where the former dominates, is the correction of purely quantum origin; at intermediate temperatures it is of quantum-thermal character, and at high temperatures purely thermal (see Sec.~\ref{sec:lpa} below). What changes across these regimes is not the status of the background field, which is $\bar x=\langle \hat x\rangle$ throughout, but the nature of the average. At high temperature, the quantum variance $\langle\hat\eta^2\rangle$ goes over to the variance of a classical stochastic variable, so that $\bar x$ may equivalently be read as the average over realizations of the classical Langevin process \eqref{eq:eom_class}, and Eq.~\eqref{eq:eom_corr} as the equation obeyed by its mean. This unity of the formalism for the quantum and classical cases also applies to the Ehrenfest discussion, simply by reinterpreting $\hat x$ as a stochastic variable and $\langle\,.\,\rangle$ a classical average over realizations.

\emph{iii. Fluctuations around the average.} Eq.~\eqref{eq:eom_corr}, which we may rewrite compactly as 
\begin{equation}\label{eq:E}
    \mathscr E[\bar x] = 0,
\end{equation}
is deterministic because it follows from the saddle-point condition \eqref{eq:saddle} of the quantum effective action
\begin{align}\label{eq:gamma_gen}
    \Gamma[\bar x, \bar x_q] &= \int dt\,\mathscr E[\bar x]\,\bar x_q\nonumber\\
    &+\frac{i}{2}\int dt\,dt'\,\bar x_q(t)[\alpha_K(t-t')+\mathscr O(\hbar)]\bar x_q(t').
\end{align}
This is what identifies $\bar x$ with the average $\langle\hat x\rangle$, making $\langle\hat\eta\rangle=0$ and $\sigma=\langle\hat\eta^2\rangle$ meaningful, and it is consistent with the Ehrenfest derivation above, where the noise likewise disappears because $\langle\hat\xi\rangle=0$. 

One may also be interested in the fluctuations $\delta x$ around the average. Their variance follows directly from the second derivative of $\Gamma$, which at $\bar x_q=0$ reads
\begin{equation}
    \Gamma^{(2)}[\bar x, \bar x_q=0] = \begin{pmatrix}
        0 & \mathcal M_\text{eff}^\dag \\ \mathcal M_\text{eff} & i\alpha_K +\mathscr O(\hbar)
    \end{pmatrix},
\end{equation}
where $\mathcal M_\text{eff} = \delta\mathscr E/\delta\bar x$. Since $i[\Gamma^{(2)}]^{-1}$ is the quantum-corrected propagator, we have
\begin{equation}\label{eq:prop}
    \langle \delta x^2\rangle = i[\Gamma^{(2)}]^{-1}_{11} = \int \frac{d\omega}{2\pi}\frac{\alpha_K(\omega)+\mathscr O(\hbar)}{|\mathcal M_\text{eff}(\omega)|^2},
\end{equation}
which reduces to $\sigma(\bar x)$ at the order retained, since both the difference between $\mathcal M_\text{eff}$ and the bare $\mathcal M$ of Eq.~\eqref{eq:M} and the correction to $\alpha_K$ discussed above are themselves $\mathscr O(\hbar)$. The corresponding dynamics is obtained by performing the HS transformation on $\Gamma$ and varying the resulting MSR action, which yields a stochastic equation for the trajectory $x_\xi=\bar x+\delta x$: $\mathscr E[x_\xi] + \xi(t)=0$. Since connected correlators are generated by the tree-level diagrams of $\Gamma$, this equation is to be used at linear order in $\delta x$ \footnote{Solving the equation $\mathscr E[x_\xi] + \xi(t)=0$ while retaining the nonlinearities would amount to computing loops with $\Gamma$, thereby overcounting the fluctuations already resummed into $\mathscr E$.},
\begin{equation}\label{eq:E_lin}
    \frac{\delta\mathscr E}{\delta\bar x}\delta x + \xi(t)=0,
\end{equation}
which reproduces Eq.~\eqref{eq:prop}, since $\langle\xi(t)\xi(t')\rangle=\alpha_K(t-t')$. At this order $\langle\delta x\rangle=0$, so that Eq.~\eqref{eq:E} for the average is unaffected, and average and fluctuations are determined consistently at the same order.

The remaining question is how to compute $\sigma(\bar x)$. Since the operator $\mathcal M$ is a nonlocal functional of $\bar x(t)$, Eq.~\eqref{eq:def_sigma} cannot be evaluated exactly. However, assuming that $\bar x(t)$ varies slowly in time, one can build a derivative expansion around a constant $\bar x$, which yields asymptotically a local expression for $\sigma(\bar x)$.

\subsection{Local potential approximation}\label{sec:lpa}

The first order in the derivative expansion is the local potential (or adiabatic) approximation, in which one treats the background field as constant. Then $\mathcal M(\omega)= m\omega^2 + i\gamma\omega-U''(\bar x)$, and $\sigma(\bar x)$ can be written exactly in terms of the digamma function $\psi(z)$. Introducing
\begin{equation}\label{eq:omegaeff}
\omega_\gamma(\bar x) = \sqrt{\frac{U''(\bar x)}{m} - \left(\frac{\gamma}{2m}\right)^2},
\end{equation}
that is the classical frequency of underdamped oscillations, and
\begin{equation}\label{eq:lambda}
    \lambda_\pm(\bar x) = \frac{\gamma}{2m}\pm i \omega_\gamma(\bar x),
\end{equation}
we obtain the expression (Appendix \ref{app:LPA})
\begin{equation}\label{eq:sigma_LPA}
    \sigma(\bar x) = \frac{k_BT}{U''(\bar x)}+\frac{\hbar}{\pi m}\frac{\psi(1+\frac{\beta\hbar\lambda_-}{2\pi})-\psi(1+\frac{\beta\hbar\lambda_+}{2\pi})}{\lambda_- - \lambda_+}.
\end{equation}
Since $\sigma(\bar x)$ depends on $\bar x$ only through $U''(\bar x)$, the correction to the EoM $\frac{1}{2}U'''(\bar x)\sigma(\bar x)$ can be written as the total derivative $U'_1(\bar x)$, with $U(\bar x)+U_1(\bar x)$ defining the effective temperature-dependent potential
\begin{align}\label{eq:ueff}
    U_{\text{eff}}(\bar x) &= U(\bar x) + U_1(\bar x)\nonumber\\
    &=U(\bar x) +\frac{1}{\beta}\ln\!\left[\frac{\beta\hbar\omega_0(\bar x)}{\Gamma(1+\frac{\beta\hbar\lambda_+}{2\pi}) \Gamma(1+\frac{\beta\hbar\lambda_-}{2\pi})}\right],
\end{align}
where $\Gamma(z)$ is the gamma function. In particular, Eq.~\eqref{eq:gamma1} can be written as
\begin{equation}\label{eq:gamma1_ueff}
    \Gamma_1 = -\int dt\,U'_1(\bar x)\,\bar x_q,
\end{equation}
so that $\Gamma$ retains the same form as $S$, with the bare potential $U$ replaced by $U_{\text{eff}}$, and Eq.~\eqref{eq:eom_corr} assumes a form analogous to Eq.~\eqref{eq:eom_class},
\begin{equation}
    m\ddot{\bar x} + \gamma\dot{\bar x} + U'_{\text{eff}}(\bar x)=0.
\end{equation}
Following Eq.~\eqref{eq:E_lin}, fluctuations around the average then satisfy
\begin{equation}\label{eq:fluc_LPA}
    m\,\delta \ddot x + \gamma\,\delta\dot x + U''_\text{eff}(\bar x)\delta x = \xi(t).
\end{equation}

Eqs.~\eqref{eq:sigma_LPA} and \eqref{eq:ueff} hold for arbitrary temperature and arbitrary damping. The reality condition for $\sigma(\bar x)$, and hence $U_1(\bar x)$, is $U''(\bar x)>0$. This is equivalent to requiring that both zeros of $\mathcal M(\omega)$, occurring at $\omega_\pm = -i\lambda_\pm$, lie in the lower half plane, i.e., the background be dynamically stable. Indeed, for $U''(\bar x)<0$ the frequency $\omega_\gamma$ becomes imaginary and $\lambda_+$ is real and negative; one zero of $\mathcal M(\omega)$ crosses into the upper half plane and the contour argument of Appendix~\ref{app:LPA} ceases to apply. The homogeneous solutions of Eq.~\eqref{eq:fluc_LPA} then grow as $e^{|\lambda_+|t}$, and no stationary fluctuation state exists. Equivalently, continuing Eq.~\eqref{eq:ueff} into that region (with the branch fixed by $U''\to U''-i0$) yields an imaginary part which, at $T=0$, reads $\mathrm{Im}\,U_1=-\frac{\hbar}{2}|\lambda_+|$; the vacuum persistence probability $|\langle 0_\text{out}|0_\text{in}\rangle|^2=|e^{-iU_\text{eff}t/\hbar}|^2=e^{2\,\mathrm{Im}\,U_1 t/\hbar}$ thus decays at the rate $|\lambda_+|$, matching the growth of fluctuations and signaling the vacuum instability. In the conservative limit, $|\lambda_+|$ reduces to the inverted-oscillator result $\sqrt{|U''(\bar x)|/m}$, while in the overdamped limit it becomes $|U''(\bar x)|/\gamma$, showing that dissipation slows the decay of the unstable vacuum. We will return to this point in Sec.~\ref{sec:conclusions}.

It is interesting to analyze the low-temperature (quantum) and the high-temperature (classical) limits of Eq.~\eqref{eq:ueff} (see Appendix \ref{app:LPA} for detailed derivations).

In the \emph{low-temperature limit},
\begin{align}
    U_\text{eff}(\bar x) &= U(\bar x)\nonumber\\
    &+ \frac{\hbar}{\pi}\left[\omega_\gamma(\bar x)\tan^{-1}\!\left(\frac{2m\omega_\gamma(\bar x)}{\gamma}\right)-\frac{\gamma}{2m}\ln\frac{\omega_0(\bar x)}{\omega_c}\right]\nonumber\\
    &-\frac{\pi\gamma(k_B T)^2}{6m\hbar\omega_0^2(\bar x)} + \mathscr O(T^4),
\end{align}
where $\omega_c$ is a reference frequency introduced for dimensional reasons and which does not appear in the EoM. The validity of the expression above extends to the overdamped regime where $\omega_\gamma$ is imaginary. The leading thermal correction is proportional to $T^2$ and thus yields an entropy linear in $T$, vanishing for $T\to 0$. Therefore, for $T \to 0$ the correction to the potential is due to purely quantum fluctuations of the system coupled to the bath. In the additional limit $\gamma \to 0$, where the system is isolated and $\omega_\gamma \to \omega_0$, we recover the conservative result $U_\text{eff}(\bar x) \simeq U(\bar x) + \frac{\hbar\omega_0(\bar x)}{2}$ \cite{furutani2022, vianello2025b, salasnich2025}, in which the correction is set by the system's own zero-point fluctuations.

In the \emph{high-temperature limit},
\begin{align}
    U_\text{eff}(\bar x) &= U(\bar x)+k_BT \ln \frac{\hbar\omega_0(\bar x)}{k_BT}\nonumber\\
    &+\frac{\hbar^2\omega_0^2(\bar x)}{24k_B T} - \frac{\zeta(3)}{8\pi^3}\frac{\hbar^3\gamma \omega_0^2(\bar x)}{m(k_B T)^2}+\mathscr O(T^{-3}).
\end{align}
The leading term is the classical free energy of harmonic fluctuations around $\bar x$, where the logarithm corresponds to the entropy contribution. The corresponding $\sigma(\bar x)$ satisfies $\frac{1}{2}U''(\bar x)\sigma(\bar x) = \frac{1}{2} k_B T$, that is the classical equipartition of energy. Thus in the high-temperature limit, the correction to the potential is due to purely (classical) thermal fluctuations of the bath. In particular, the whole $\gamma$ dependence of $\sigma(\bar x)$ is of quantum origin, as is already clear from Eq.~\eqref{eq:sigma_LPA}.

In the following we will focus on the \emph{weak damping regime}, where we approximate $\lambda_\pm \simeq \pm i\omega_\gamma$ \footnote{With respect to the approximation $\lambda_\pm \simeq \pm i\omega_0$, which would correspond to the strictly undamped case and is discussed in Appendix \ref{app:LPA}, here we are retaining the exact imaginary part of $\lambda_\pm$, which corresponds to keeping the exact real part of the zeros of the fluctuation operator $\mathcal M(\omega)$, while discarding the linewidth. Consistently, the factor $\omega_0=\sqrt{\lambda_+\lambda_-}$ in Eq.~\eqref{eq:ueff} is also replaced by $\omega_\gamma$.}. This yields
\begin{equation}\label{eq:Veff}
U_{\text{eff}}(\bar x) \simeq U(\bar x) + \frac{\hbar\omega_\gamma(\bar x)}{2} + \frac{1}{\beta}\ln\left(1-e^{-\beta\hbar\omega_\gamma(\bar x)}\right),
\end{equation}
showing that---analogously to the conservative case, which is recovered in the limit $\gamma\to 0$ as $\omega_\gamma \to \omega_0$ \cite{salasnich2025}---the correction is set by the energy of quantum-thermal fluctuations at the damped frequency $\omega_\gamma(\bar x)$, which is given by the zero-point contribution plus the Bose occupation.

\subsection{Derivative expansion}\label{sec:derexp}

Going beyond the local potential approximation is in general a difficult task. If we restrict to the regime of low temperature and weak damping, which is the relevant one for the applications to Josephson junction we will discuss in Sec.~\ref{sec:applications}, we can instead make rapid progress. In fact, Eq.~\eqref{eq:gamma1_ueff} simplifies to
\begin{equation}
    \Gamma_1 \simeq -\int dt\,\frac{\partial}{\partial\bar x}\frac{\hbar\omega_\gamma(\bar x)}{2}\,\bar x_q,
\end{equation}
meaning that quantum corrections are governed by the zero-point energy $\frac{\hbar}{2}\omega_\gamma(\bar x)$. Consistency with the conservative limit \cite{furutani2022, vianello2025b} then requires that, in this regime, the quantum correction \eqref{eq:gamma1} matches
\begin{equation}\label{eq:gamma1_ref}
    \Gamma_1 = \frac{i\hbar}{2}\int dt\,\frac{\delta}{\delta\bar x}\int\frac{d\omega}{2\pi}\ln\left[\omega^2-\omega_\gamma^2(\bar x)\right] \bar x_q
\end{equation}
at all orders in the derivative expansion, where the functional derivative is 
\begin{equation}\label{eq:funcder}
    \frac{\delta}{\delta\bar x}=\frac{\partial}{\partial\bar{x}}-\frac{d}{dt}\frac{\partial}{\partial \dot{\bar x}}.
\end{equation}
Using the derivative expansion of ${\int \frac{d\omega}{2\pi}\ln[\omega^2-\omega_\gamma^2(\bar x)]}$ known from the conservative case \cite{kleinert_book, jackiw1974, iliopoulos1975, furutani2022, vianello2025b, fraser1985, cametti2000}, we obtain
\begin{equation}\label{eq:devexp}
    \Gamma_1 = \frac{\hbar}{2}\int dt\,\frac{\delta}{\delta\bar x}\biggl[-\omega_\gamma(\bar x) +\frac{[\partial_{\bar x} \omega_\gamma^2(\bar x)]^2}{32\omega_\gamma^5(\bar x)}\dot{\bar x}^2+\mathscr O(\dot{\bar x}^4, \ddot{\bar x})\biggr] \bar x_q.
\end{equation}
The absence of odd powers of $\dot{\bar x}$ in the derivative expansion is a direct consequence of the invariance of Eq.~\eqref{eq:gamma1_ref} under time reversal. The term proportional to $\dot{\bar x}^2$ produces a renormalization of the mass $m$, which is replaced by the effective mass
\begin{equation}\label{eq:meff}
    m_\text{eff}(\bar x) = m + \frac{\hbar}{32}\frac{[\partial_{\bar x} \omega_\gamma^2(\bar x)]^2}{\omega_\gamma^5(\bar x)}.
\end{equation}
We thus conclude that the quantum-corrected EoM at low temperature and weak damping (at one loop and second order in the derivative expansion) is
\begin{equation}\label{eq:motion}
    m_\text{eff}(\bar x)\ddot{\bar x} + \frac{m'_\text{eff}(\bar x)}{2}\dot{\bar x}^2+ \gamma \dot{\bar x} + U_\text{eff}'(\bar x) = 0,
\end{equation}
with $U_\text{eff}(\bar x) = U(\bar x) + \frac{\hbar}{2}\omega_\gamma(\bar x)$. An alternative derivation of this result, based on a generalized Hamilton principle, is discussed in Appendix \ref{app:C}. Following Eq.~\eqref{eq:E_lin}, fluctuations around the average $\bar x$ then satisfy
\begin{align}\label{eq:fluctuations}
    &m_\text{eff}(\bar x)\delta\ddot x + \left[\gamma + m'_\text{eff}(\bar x)\dot{\bar x}\right]\delta\dot x \nonumber\\
    &+ \left[U''_\text{eff}(\bar x) + m'_\text{eff}(\bar x)\ddot{\bar x} + \frac{m''_\text{eff}(\bar x)}{2}\dot{\bar x}^2\right]\delta x = \xi(t).
\end{align}

\subsection{Phase space formulation}\label{sec:phasespace}

With a view toward subsequent applications to bosonic Josephson junctions, it is useful to extend the preceding results to the case in which the conservative action is formulated in phase space, namely 
\begin{equation}
S_0[x,p] = \int dt \left[p\dot x-H(x,p)\right],
\end{equation}
where $H(x,p)$ is the Hamiltonian. Even in the absence of dissipation, this is an interesting generalization of the method presented in Ref.~\cite{vianello2025b}. We assume that the dissipative part of the Keldysh action retains the structure given in Eq.~\eqref{eq:acK_diss}, so that the action analogous to \eqref{eq:keldysh_markov} reads
\begin{align}\label{eq:acps}
    S &= \int dt \left[\left(\dot x - \frac{\partial H}{\partial p}\right)p_q-\left(\dot p+\frac{\partial H}{\partial x}+\gamma\dot x\right)x_q\right]\nonumber\\
    &+\frac{i}{2}\int dt\,dt'x_q(t)\alpha_K(t-t')x_q(t') + \mathscr O(x_q^3, p_q^3).
\end{align}
Expanding $x$, $p$, $x_q$, and $p_q$ around background fields, $x=\bar x+\eta$, $p = \bar p + \rho$, $x_q=\bar x_q+\eta_q$, and $p_q = \bar p_q+\rho_q$, we obtain, to second order in the fluctuations $\underline{\bm\eta} = (\eta, \rho, \eta_q, \rho_q)$,
\begin{equation}
    S \simeq S[\bar x, \bar p, \bar x_q, \bar p_q] + \frac{1}{2}\int dt\,dt'\underline{\bm \eta}^T(t)\bm{\mathcal F}(t-t')\underline{\bm \eta}(t'),
\end{equation}
where
\begin{gather}
    \bm{\mathcal F} = \begin{pmatrix} -(\bar x_q \partial_{\bar x}+\bar p_q\partial_{\bar p})\mathbf H & \bm{\mathcal M}^\dag \\ \bm{\mathcal M} & i\mathbf D\alpha_K\end{pmatrix} + \mathscr O(\bar x_q^2, \bar p_q^2),\\
    \bm{\mathcal M} = \delta(t-t')[\mathbf J \partial_t -\gamma \mathbf D \partial_t - \mathbf H(\bar x,\bar p)],
\end{gather}
with $\mathbf{J} = \left(\begin{smallmatrix}0&-1\\1&0\end{smallmatrix}\right)$, $\mathbf D = \left(\begin{smallmatrix}1&0\\0&0\end{smallmatrix}\right)$, and
\begin{equation}
    \mathbf H = \begin{pmatrix}
        H_{\bar x \bar x} & H_{\bar x \bar p} \\ H_{\bar p \bar x} & H_{\bar p \bar p}
    \end{pmatrix}
\end{equation}
the Hessian of the Hamiltonian evaluated on $(\bar x, \bar p)$. The structural similarity with Eqs.~\eqref{eq:F} and \eqref{eq:M} is evident. In the local potential approximation, $\Gamma$ thus retains the same form as the action \eqref{eq:acps}, with the Hamiltonian $H$ replaced by the effective Hamiltonian
\begin{equation}
    H_\text{eff}(\bar x, \bar p) = H(\bar x, \bar p) + U_1(\bar x, \bar p),
\end{equation}
where $U_1$ is given by Eq.~\eqref{eq:ueff} with $\omega_0$ replaced by $\sqrt{\det \mathbf H(\bar x, \bar p)}$ and $\lambda_\pm = \frac{\gamma H_{\bar p \bar p}}{2} \pm i\omega_\gamma(\bar x, \bar p)$, where
\begin{equation}\label{eq:omegaeffH}
\omega_\gamma(\bar x, \bar p) = \sqrt{\det \mathbf H(\bar x, \bar p) - \left(\frac{\gamma H_{\bar p \bar p}}{2}\right)^2}.
\end{equation}

In particular, at low temperature and weak damping, the quantum correction is given by
\begin{equation}
    \Gamma_1 \simeq -\frac{\hbar}{2}\int dt \left(\bar x_q\frac{\partial}{\partial\bar x} + \bar p_q\frac{\partial}{\partial \bar p}\right)\omega_\gamma(\bar x, \bar p).
\end{equation}
Since $(\partial_t \omega^2_\gamma)^2 = (\partial_{\bar x}\omega^2_\gamma)^2\dot{\bar x}^2 + (\partial_{\bar p}\omega^2_\gamma)^2\dot{\bar p}^2 + 2 \partial_{\bar p} \omega^2_\gamma \partial_{\bar x}\omega^2_\gamma \dot{\bar p}\dot{\bar x}$, the derivative expansion in Eq.~\eqref{eq:devexp} generalizes to
\begin{align}
    \Gamma_1 = \frac{\hbar}{2}\int dt&\left(\bar x_q\frac{\delta}{\delta\bar x}+\bar p_q\frac{\delta}{\delta\bar p}\right)\!\biggl[-\omega_\gamma(\bar x, \bar p)\nonumber\\
    &+ \frac{Z_{\bar x\bar x}}{2}\dot{\bar x}^2 + \frac{Z_{\bar p \bar p}}{2}\dot{\bar p}^2 + Z_{\bar p \bar x}\dot{\bar p}\dot{\bar x}+\cdots\biggr],
\end{align}
where $Z_{\bar x \bar y} = \partial_{\bar x} \omega_\gamma^2(\bar x, \bar p) \partial_{\bar y}\omega_\gamma^2(\bar x, \bar p)/16 \omega_\gamma^5(\bar x, \bar p)$ and the functional derivative $\delta/\delta\bar p$ is defined analogously to Eq.~\eqref{eq:funcder}. It is clear that, in general, these derivative corrections deform the canonical structure. Defining the derivative deformation
\begin{equation}\label{eq:Q}
    Q[\bar x, \bar p] = \frac{Z_{\bar x\bar x}}{2}\dot{\bar x}^2 + \frac{Z_{\bar p \bar p}}{2}\dot{\bar p}^2 + Z_{\bar p \bar x}\dot{\bar p}\dot{\bar x},
\end{equation}
the quantum effective action reads
\begin{align}
    \Gamma &= \int dt \left[\left(\dot{\bar x} - \frac{\partial H_\text{eff}}{\partial \bar p}\right)\bar p_q-\left(\dot {\bar p}+\frac{\partial H_\text{eff}}{\partial \bar x}+\gamma\dot{\bar x}\right)\bar x_q\right]\nonumber\\
    &+ \frac{\hbar}{2}\int dt \left(\bar x_q\frac{\delta}{\delta\bar x}+\bar p_q\frac{\delta}{\delta\bar p}\right)Q[\bar x, \bar p]\nonumber\\
    &+\frac{i}{2}\int dt\,dt'\,\bar x_q(t)\alpha_K(t-t')\bar x_q(t').
\end{align}
Varying $\Gamma$ then yields the quantum-corrected EoMs
\begin{subequations}\label{eq:effH}
\begin{align}
    \dot{\bar x} - \frac{\partial H_\text{eff}}{\partial \bar p} + \frac{\hbar}{2}\frac{\delta Q}{\delta\bar p} &= 0,\label{eq:effH1}\\
    \dot {\bar p}+\frac{\partial H_\text{eff}}{\partial \bar x}+\gamma\dot{\bar x}- \frac{\hbar}{2}\frac{\delta Q}{\delta\bar x} &= 0.\label{eq:effH2}
\end{align}
\end{subequations}
This result simplifies substantially, and matches Eq.~\eqref{eq:motion}, when the Hamiltonian takes the form $H(x, p) = p^2/2m+U(x)$. In fact, in this case $H_{\bar x \bar x}=U''(\bar x)$, $H_{\bar p \bar p}=1/m$, and $H_{\bar x \bar p}=0$, therefore Eq.~\eqref{eq:omegaeffH} is equivalent to Eq.~\eqref{eq:omegaeff} and independent of $\bar p$. It follows that $H_\text{eff}(\bar x, \bar p) = \bar p^2/2m+U_\text{eff}(\bar x)$, with $U_\text{eff}$ given by Eq.~\eqref{eq:Veff} (evaluated at $T=0$), and $Z_{\bar p \bar p} = Z_{\bar p \bar x}=0$. Eq.~\eqref{eq:effH1} then yields $\bar p = m \dot{\bar x}$, which substituted into Eq.~\eqref{eq:effH2} gives exactly Eq.~\eqref{eq:motion} with $m_\text{eff}(\bar x)=m+\hbar Z_{\bar x \bar x}/2$, in agreement with Eq.~\eqref{eq:meff}.

\section{Application to Josephson junctions}\label{sec:applications}

In this section, we discuss two applications of the general results derived in Sec.~\ref{sec:theory} to superconducting and superfluid Josephson junctions.

\subsection{Resistively and capacitively shunted superconducting junction}\label{sec:rcsj}

In the RCSJ model \cite{barone_book, koch1980, koch1981, koch1982, larkin_book, furutani2021} for a superconducting circuit with capacitance $C$, shunt resistance $R$, critical current $I_c$, and external current $I$, the superconducting phase $\phi$ obeys the equation
\begin{equation}\label{eq:rcsj}
\frac{\hbar^2 C}{4e^2}\ddot\phi + \frac{\hbar^2}{4e^2R}\dot\phi + U'(\phi) = \xi(t),
\end{equation}
with the washboard potential
\begin{equation}
    U(\phi) = -\frac{\hbar}{2e}(I\phi + I_c \cos\phi),
\end{equation}
where $\xi(t)$ is a current noise originating from the shunt resistor. Eq.~\eqref{eq:rcsj} is of the form of Eq.~\eqref{eq:eom_class} upon the identifications $x\to\phi$, $m \to \hbar^2 C/4e^2$, and $\gamma \to \hbar^2/4e^2R$. In the low-temperature and weak-damping regime, applying Eq.~\eqref{eq:Veff} evaluated at $T=0$ we thus get the effective potential
\begin{equation}
    U_\text{eff}(\bar \phi) = U(\bar \phi) + \frac{\hbar}{2}\sqrt{\Omega_J^2\cos \bar\phi - \Omega_{RC}^2},
\end{equation}
where we have introduced
\begin{equation}
    \Omega_J = \sqrt{\frac{2eI_c}{\hbar C}}, \qquad \Omega_{RC} = \frac{1}{2RC},
\end{equation}
with $\Omega_J$ being the classical Josephson frequency. The effective mass, which results in an effective capacitance, is instead given by [Eq.~\eqref{eq:meff}]
\begin{equation}
    C_\text{eff}(\bar \phi) = C + \frac{e^2}{8\hbar}\frac{\Omega_J^4 \sin^2\bar\phi}{(\Omega_J^2\cos\bar\phi-\Omega_{RC}^2)^{5/2}}.
\end{equation}
The quantum-corrected RCSJ equation is then [Eq.~\eqref{eq:motion}]
\begin{equation}\label{eq:rcsj_corr}
    \frac{\hbar^2 C_\text{eff}(\bar\phi)}{4e^2}\ddot{\bar \phi} + \frac{\hbar^2C'_\text{eff}(\bar\phi)}{8e^2}\dot{\bar \phi}^2 +\frac{\hbar^2}{4e^2R}\dot{\bar \phi} + U_\text{eff}'(\bar\phi) = 0,
\end{equation}
while fluctuations $\delta\phi$ around the average $\bar\phi$ satisfy [Eq.~\eqref{eq:fluctuations}]
\begin{align}
    &\frac{\hbar^2 C_\text{eff}(\bar\phi)}{4e^2}\delta\ddot\phi + \frac{\hbar^2}{4e^2}\!\left(\frac{1}{R}+C'_\text{eff}(\bar\phi)\dot{\bar\phi}\right)\delta\dot\phi\nonumber\\
    &+ \!\left[U_\text{eff}''(\bar\phi) + \frac{\hbar^2}{4e^2}\!\left(C'_\text{eff}(\bar\phi)\ddot{\bar\phi} + \frac{C''_\text{eff}(\bar\phi)}{2}\dot{\bar\phi}^2\right)\!\right]\delta\phi = \xi(t).
\end{align}
The influence of noise in the last equation may then be analyzed by employing the same methodology as in Refs.~\cite{furutani2021, brandt2010}. The essential difference, however, is that the underlying deterministic dynamics now intrinsically incorporates quantum fluctuations of the macroscopic degree of freedom.

Let us consider a simple instance of such dynamics. In the absence of external current ($I=0$), the effective potential has a minimum in $\bar\phi=0$, where $C_\text{eff}(0)=C$. The linearized EoM around this minimum is therefore $\ddot{\bar \phi} + 2\Omega_{RC}\dot{\bar\phi} + \tilde\Omega_0^2\bar\phi = 0$, where $\tilde\Omega_0^2=4e^2U''_\text{eff}(0)/\hbar^2 C = \Omega_J^2[1-(e^2/\hbar C) (\Omega_J^2-\Omega_{RC}^2)^{-1/2}]$. Its solution takes the form
\begin{equation}
    \bar\phi(t) = A\,e^{-\Omega_{RC} t} \cos(\tilde \Omega_d t + c),
\end{equation}
where $A$ and $c$ are fixed by the initial conditions, and the oscillation frequency is $\tilde\Omega_d = \sqrt{\tilde\Omega_0^2-\Omega_{RC}^2}$, namely
\begin{equation}\label{eq:omegarcsj}
    \tilde \Omega_d = \sqrt{\Omega_J^2-\Omega_{RC}^2}\sqrt{1-\frac{(E_C/2E_J)^{1/2}}{[1-(\Omega_{RC}/\Omega_J)^2]^{3/2}}},
\end{equation}
with $E_C = e^2/2C$ and $E_J = \hbar I_c/2e$. In the product on the right-hand side, the first term, $\Omega_d = \sqrt{\Omega_J^2-\Omega_{RC}^2}$, corresponds to the classical frequency of underdamped oscillations, while the second term encodes the quantum correction. In the absence of damping ($\Omega_{RC}=0$), Eq.~\eqref{eq:omegarcsj} simplifies to $\tilde \Omega_0 = \Omega_J\sqrt{1-(E_C/2E_J)^{1/2}}$, in agreement with the expression reported in Ref.~\cite{furutani2022}. Unlike the oscillation frequency, the decay rate $\Omega_{RC}$ acquires no quantum correction, at least in the linear regime around $\bar\phi=0$.

The RCSJ model has several experimental realizations. In small Al/AlOx/Al junctions used in superconducting qubits, one typically has $C \sim 10-100 \text{ fF}$, $R \sim 1-10 \text{ k$\Omega$}$, and $I_c \sim 10-1000 \text{ nA}$. These parameters place the junction deeply in the underdamped regime, $\Omega_{RC} \ll \Omega_J$. Taking for instance $C= 70 \text{ fF}$, $R = 5 \text{ k$\Omega$}$, and $I_c = 20 \text{ nA}$, we obtain $\Omega_J \simeq 29.5 \text{ GHz}$ and $\Omega_{RC} \simeq 1.43 \text{ GHz}$. Eq.~\eqref{eq:omegarcsj} then yields $\tilde \Omega_d \simeq \tilde \Omega_0 \simeq 0.94\Omega_J$, so quantum corrections produce a $\sim 6\%$ reduction of the Josephson frequency. In contrast, shunted Nb/AlOx/Nb junctions used in classical superconducting electronics (such as RSFQ logic or voltage standards) typically have $C \sim 10-1000 \text{ fF}$, $R \sim 1-100 \text{ $\Omega$}$, and $I_c \sim 10-1000 \text{ $\mu$A}$. Taking for instance $C= 10 \text{ fF}$, $R = 20 \text{ $\Omega$}$, and $I_c = 100 \text{ $\mu$A}$, we obtain $\Omega_J \simeq 5.51 \text{ THz}$ and $\Omega_{RC} = 2.50 \text{ THz}$, hence $\Omega_{RC}/\Omega_J \simeq 0.45$, implying sizable damping. Eq.~\eqref{eq:omegarcsj} then yields $\Omega_d = \sqrt{\Omega_J^2-\Omega_{RC}^2}\simeq 4.91 \text{ THz}$ and $\tilde\Omega_d\simeq 0.997\Omega_d$, indicating that quantum corrections reduce the frequency by $\sim 0.3\%$.

Although Eq.~\eqref{eq:omegarcsj} has been obtained in the low-temperature regime, its extension to finite temperature is immediate. At the equilibrium configuration considered here the derivative corrections vanish, so that the dynamics is governed by the local potential approximation alone, which by Eq.~\eqref{eq:Veff} is exact in $T$. Since the first derivative of $\omega_\gamma$ also vanishes at that point, only the derivative $\partial U_1/\partial\omega_\gamma=\frac{\hbar}{2}\coth(\beta\hbar\omega_\gamma/2)$ enters; consequently, Eq.~\eqref{eq:omegarcsj} extends to finite temperature through the single substitution $\sqrt{E_C/2E_J}\to \sqrt{E_C/2E_J} \coth(\beta\hbar\Omega_d/2)$, which amounts to replacing the zero-point energy by the full energy of quantum-thermal fluctuations at the underdamped frequency. The corrections are thus not suppressed at high temperature: they cross over, at 
\begin{equation}
    T^* = \frac{\hbar\Omega_d}{2k_B},
\end{equation}
from quantum to thermal, growing linearly in $T$. Al/AlOx/Al junctions have a typical operating temperature $T \sim 10 \text{ mK}$, while $\Omega_d \simeq 29.5 \text{ GHz}$ gives $T^* \simeq 110 \text{ mK}$. Similarly, Nb/AlOx/Nb junctions typically have $T \sim 4 \text{ K}$, while $\Omega_d\simeq 4.91 \text{ THz}$ yields $T^* \simeq 19 \text{ K}$. Hence both the superconducting examples above are deeply in the quantum regime, where corrections are controlled by $\sqrt{E_C/2E_J}$. This is of the order of $10^{-1}$ and $10^{-3}$ for the two examples, respectively, well inside the regime of validity of the one-loop expansion.

\subsection{Elongated bosonic junction}\label{sec:ebj}

In an elongated bosonic junction realized by two one-dimensional Bose gases coupled through a point-like tunneling barrier, dissipation originates from the intrinsic interaction between the collective Josephson mode and the bath of elementary quasiparticle excitations supported by the condensates (see, e.g., Refs.~\cite{pigneur2018, polo2018, binanti2021}). As shown in detail in Appendix \ref{app:B}, integrating out the bath degrees of freedom, and using for convenience dimensionless units for time and energy \cite{smerzi1997}, yields an effective action for the Josephson mode of the form given in Eq.~\eqref{eq:acps}, with the identification $x\to \phi$, the relative phase, and $p \to z$, the population imbalance between the two condensates. The Hamiltonian is
\begin{equation}\label{eq:V2}
H(\phi,z) = \frac{\Lambda}{2}z^2 - \sqrt{1-z^2}\cos\phi,
\end{equation}
where $\Lambda>0$ encodes the boson-boson interaction. In the usual low-temperature and weak-damping regime, Eq.~\eqref{eq:omegaeffH} gives
\begin{align}\label{eq:omegaeffbos}
\omega_\gamma(\bar\phi, \bar z) &=\sqrt{\omega_0^2(\bar\phi, \bar z)-\frac{\gamma^2}{4}\left[\Lambda + \frac{\cos\bar\phi}{(1-\bar z^2)^{3/2}}\right]^2},
\end{align}
where
\begin{align}
    \omega_0^2(\bar\phi, \bar z) &= \Lambda \sqrt{1- \bar z^2}\cos \bar \phi +\frac{\cos^2 \bar \phi - \bar z^2 \sin^2 \bar\phi}{1-\bar z^2}.
\end{align}
The resulting effective Hamiltonian is
\begin{equation}\label{eq:Veff2}
H_\mathrm{eff}(\bar\phi,\bar z) = H(\bar\phi,\bar z) + \frac{\hbar\omega_\gamma(\bar\phi,\bar z)}{2}.
\end{equation}
Note that since we are working with adimensional time and energy, in this section $\hbar$ is also adimensional, and it is given by $\hbar=2/N$. The Hamiltonian has a stable equilibrium in $(\bar\phi, \bar z) = \bm 0$, where, since $\partial_{\bar\phi}\omega_\gamma^2(\bm 0) = \partial_{\bar z}\omega_\gamma^2(\bm 0)=0$, the derivative corrections in Eq.~\eqref{eq:Q} and in the corresponding linearized EoMs \eqref{eq:effH} also vanish. The latter thus take the simple form
\begin{subequations}\label{eq:lin}
\begin{align}
    \dot{\bar\phi} -\partial_{\bar z}^2 H_\text{eff}(\bm 0)\bar z &= 0,\label{eq:lin1}\\
    \dot{\bar z} + \partial_{\bar\phi}^2 H_\text{eff}(\bm 0)\bar\phi + \gamma\dot{\bar\phi} &= 0\label{eq:lin2}.
\end{align}
\end{subequations}
Defining
\begin{subequations}
\begin{align}
    \Omega_J &= \omega_0(\bm 0)=\sqrt{\Lambda+1},\\
    \Omega_d &= \omega_\gamma(\bm 0) = \Omega_J \sqrt{1-\left(\frac{\gamma\Omega_J}{2}\right)^2},
\end{align}
\end{subequations}
as the classical Josephson and underdamped frequencies, respectively, substituting Eq.~\eqref{eq:lin2} into Eq.~\eqref{eq:lin1} we obtain a single second-order equation for $\bar\phi(t)$:
\begin{equation}\label{eq:damped_osc}
    \ddot{\bar\phi} + \tilde \Gamma \dot{\bar \phi} + \tilde \Omega_0^2 \bar \phi = 0,
\end{equation}
where
\begin{subequations}\label{eq:corrbos1}
\begin{align}
    \tilde \Omega_0^2 &= \Omega_J^2 + \frac{\hbar}{4\Omega_d}\left[3-\Omega_J^2 \left(2+\frac{3\gamma^2}{2}\right)-\Omega_J^4\left(1-\frac{\gamma^2}{2}\right)\right],\\
    \tilde \Gamma &= \gamma \left\{\Omega_J^2 +\frac{\hbar}{4\Omega_d}\left[3 - \Omega_J^2\left(1+\frac{3\gamma^2}{2}\right)\right]\right\}.
\end{align}
\end{subequations}  
The dynamics of $\bar z$ is then completely determined by that of $\bar\phi$ via the inverse of Eq.~\eqref{eq:lin1}, $\bar z = [\partial_{\bar z}^2H_\text{eff}(\bm 0)]^{-1}\dot{\bar\phi}$.

We notice that in this case, due to the coupling between $\phi$ and $z$, also the damping coefficient acquires a quantum correction. The solution of Eq.~\eqref{eq:damped_osc} is
\begin{equation}
    \bar\phi(t) = A\,e^{-\tilde\Gamma t/2} \cos(\tilde \Omega_d t + c),
\end{equation}
where $A$ and $c$ are fixed by the initial conditions, and the oscillation frequency is $\tilde\Omega_d=\sqrt{\tilde \Omega_0^2- (\tilde \Gamma/2)^2}$, namely
\begin{align}
    \tilde\Omega_d &= \Omega_d\sqrt{1 \!+\! \frac{\hbar}{4\Omega_d^3}\!\left[3 \!-\! \Omega_J^2\!\left(2 \!+\! 3\gamma^2\right) \!-\! \Omega_J^4\!\left(1 \!-\! \gamma^2 \!-\! \frac{3\gamma^4}{4}\right)\!\right]}.
\end{align}
In the absence of dissipation ($\gamma=0$), this reduces to
\begin{equation}\label{eq:corrbos2}
    \tilde\Omega_0 = \Omega_J \sqrt{1+\frac{\hbar}{4\Omega_J^3}\left(3-2\Omega_J^2-\Omega_J^4\right)}.
  \end{equation}
It is clear from Eq.~\eqref{eq:lin1} that the population imbalance has a dynamics of the same form, oscillating with frequency $\tilde \Omega_d$ while its amplitude decays as $\exp(-\tilde \Gamma t/2)$.

\begin{figure}
    \centering
    \includegraphics[width=\linewidth]{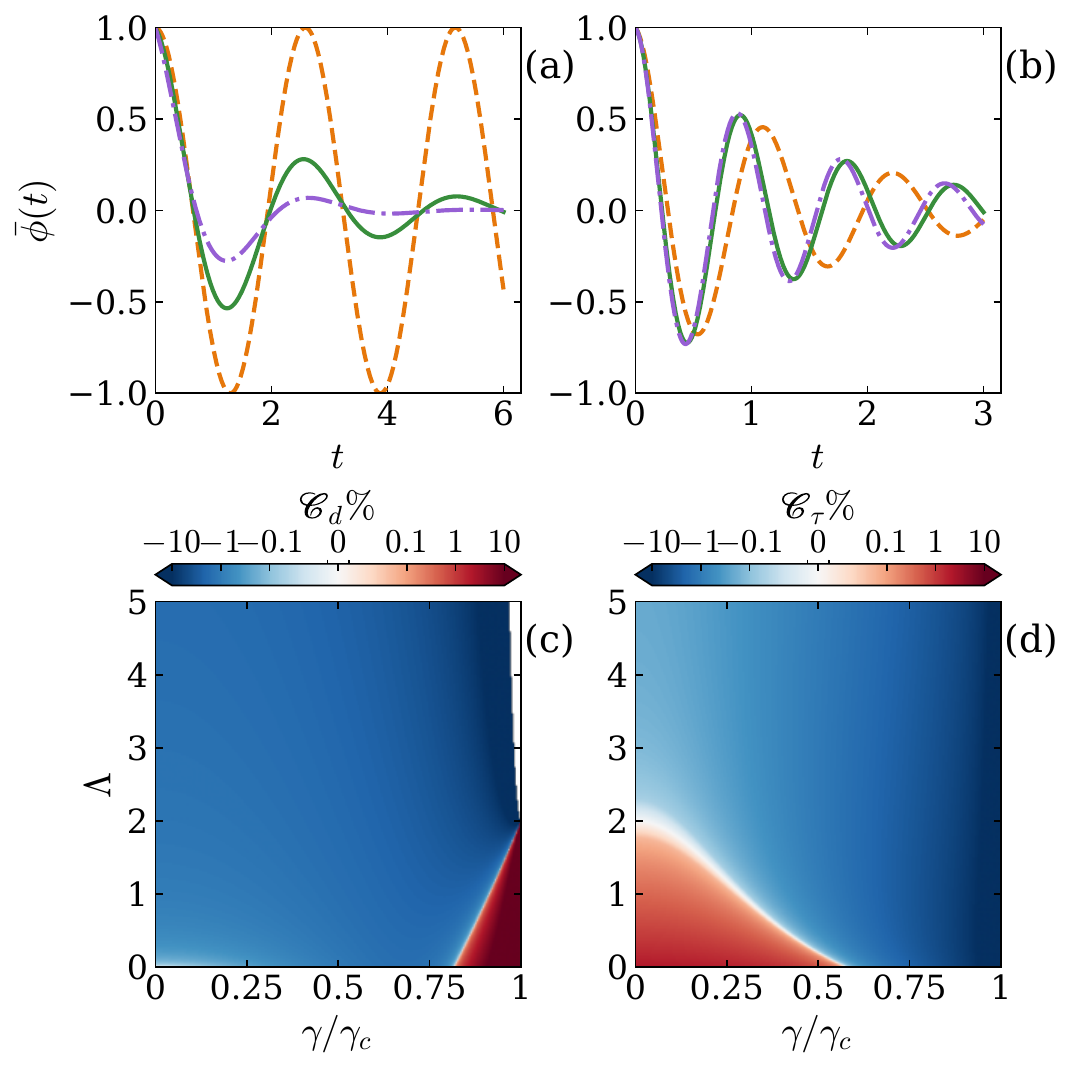}
    \caption{Zero-temperature quantum corrections to the dynamics of the bosonic junction. (a) Phase dynamics for $N=100$, $\Lambda=5$, and  $\gamma=0$ (dashed orange), $\gamma=0.2\gamma_c$ (solid green), and $\gamma=0.4\gamma_c$ (dash-dotted violet), where $\gamma_c=2/\Omega_J$.  (b) Phase dynamics for $\Lambda=50$, $\gamma=0.1\gamma_c$ and  $N=10$ (dashed orange), $N=50$ (solid green), and $N=200$ (dash-dotted violet). (c) Quantum correction to the Josephson frequency, $\mathscr{C}_d=\tilde\Omega_d/\Omega_d-1$, for $N=50$, as a function of $\gamma/\gamma_c$ and $\Lambda$. (d) Quantum correction to the damping coefficient, $\mathscr{C}_\tau=\tilde\Gamma/\gamma\Omega_J^2-1$, for $N=50$, as a function of $\gamma/\gamma_c$ and $\Lambda$.}
    \label{fig1}
\end{figure}

These quantum corrections can easily reach a few percent and, in the presence of dissipation, are generally larger than in the undamped case (Fig.~\ref{fig1}). For instance, fixing $\Lambda=50$ and $\gamma=0.2\gamma_c$, where $\gamma_c=2/\Omega_J$, we find that $\mathscr C = \tilde\Omega_0/\Omega_J-1$ amounts to $-6.4\%$ for $N=30$, $-1.9\%$ for $N=100$, and $-0.9\%$ for $N=200$, while $\mathscr C_d = \tilde\Omega_d/\Omega_d-1$ reaches $-6.8\%$, $-2.0\%$, and $-1.0\%$ for the same value of $N$, respectively. Under the same conditions, the quantum correction to the damping coefficient, $\mathscr C_\tau = \tilde\Gamma/\gamma\Omega_J^2-1$, is $-0.23\%$ for $N=30$, $-0.07\%$ for $N=100$, and $-0.03\%$ for $N=200$.

The same argument given at the end of Sec.~\ref{sec:rcsj} applies here, and Eqs.~\eqref{eq:corrbos1}-\eqref{eq:corrbos2} extend to finite temperature through the substitution $\hbar\to\hbar\coth(T^*/T)$, the three corrections $\mathscr C$, $\mathscr C_d$ and $\mathscr C_\tau$ being all multiplied by the same factor. Since here both $\hbar$ and $\Omega_d$ are adimensional, and times are measured in units of $\hbar/2J$, with $J$ the tunneling energy (see Appendix \ref{app:B} and Ref.~\cite{smerzi1997}), the physical energy quantum associated with $\Omega_d$ is $2 J\Omega_d$, so that
\begin{equation}
    T^*=\frac{J \Omega_d}{k_B}.
\end{equation}
The quantum-thermal corrections are thus governed by $\hbar\coth(T^*/T) \simeq 2k_BT/E_J\Omega_d$ for $T\gg T^*$, with $E_J=JN$ the total Josephson coupling energy. The relevant parameter at high temperature is then $k_BT/E_J$, and the one-loop treatment requires $k_BT\ll E_J$, which is also the condition for the relative phase to remain locked.

Bosonic junctions realized with ultracold atoms may operate on either side of the crossover, depending on the barrier height and on whether the Josephson mode has equilibrated with the thermal cloud; in the thermal regime the fluctuations of the relative phase are of classical origin, and their measurement provides a primary thermometer for the ultracold gas \cite{Gati2006}. Room-temperature magnon condensates \cite{Demokritov2006, Kreil2021, Nakata2014, Nakata2024b} lie instead deep in the thermal regime, but there the corrections remain negligible because the number of magnons is macroscopically large and $k_BT/E_J$ correspondingly small. Our treatment thus quantifies why purely mean-field, formally zero-temperature descriptions are usually accurate for such systems, and indicates where they would cease to be, namely at small $N$ or small tunneling energy.

\section{Conclusions and perspectives}\label{sec:conclusions}

In summary, we have applied the quantum effective action approach within the Schwinger-Keldysh formalism to derive quantum-thermal corrections to the semiclassical dynamics of a dissipative macroscopic quantum system. Since our framework is based on an effective action, it provides a general description that is largely independent of the microscopic details of either the system or the environment, making it widely applicable. We have shown that, at one loop, corrections are encoded in a non-local functional $\sigma(\bar x)$, representing the variance of quantum-thermal fluctuations of the system coupled to the environment and depending parametrically on the temperature $T$ and the damping coefficient $\gamma$. This result is consistent with the Ehrenfest equation derived from the quantum Langevin equation. Within the local potential approximation, we have computed $\sigma(\bar x)$ exactly for arbitrary temperature and damping, showing that in the weak-damping regime it generates a correction to the system potential given by the energy of fluctuations evaluated at the classical underdamped frequency. In the additional low-temperature limit, where the zero-point energy dominates, the derivative expansion of $\sigma(\bar x)$ can be extended to higher order analogously to the conservative case, yielding quantum corrections not only to the potential but also to the effective mass. We have considered both situations in which the conservative dynamics is formulated in configuration space and in phase space, as naturally realized in the RCSJ model and in the elongated bosonic junction, respectively. For both examples, we have explicitly computed the quantum corrections to the oscillation frequency and to the decay rate of phase oscillations, showing that they can reach the percent level under realistic conditions. 

The quantum effective action framework also accommodates vacuum-instability phenomena, in which fluctuations around the background play an essential role. Two such effects are already contained in our equations. The first is the decay of a locally unstable background, which we mentioned in Sec.~\ref{sec:lpa}. For $U''(\bar x)<0$, at zero temperature $U_1(\bar x)$ develops an imaginary part $\text{Im}\,U_1=-\frac{\hbar}{2}|\lambda_+|$, implying the decay of the vacuum persistence probability at the rate $|\lambda_+|=\sqrt{|U''(\bar x)|/m+\left(\gamma/2m\right)^2}-\gamma/2m$. At finite temperature, the analytic continuation to the region where $U''(\bar x)<0$ is further constrained. Since $\lambda_+<0$, the gamma function $\Gamma(1+\beta\hbar\lambda_+/2\pi)$ has poles at $T=T_0/n$ for $n=1,2,\dots$, where
\begin{align}
    T_0 &= \frac{\hbar|\lambda_+|}{2\pi k_B}
\end{align}
coincides with the crossover temperature between thermal activation and macroscopic quantum tunneling in dissipative escape \cite{Grabert1984, hanggi1990, weiss_book}. Equivalently, noting that Eq.~\eqref{eq:ueff} may be written as the Matsubara sum (Appendix \ref{app:LPA})
\begin{align}\label{eq:ueff_matsubara}
    U_\text{eff}(\bar x) &= U(\bar x)+k_BT\ln \frac{\hbar\omega_0(\bar x)}{k_B T} \nonumber\\
    &+ k_BT\sum_{n\ge 1}\ln\left[\nu_n^2+\frac{\gamma}{m}|\nu_n|+\omega_0^2(\bar x)\right],
\end{align}
the temperature $T_0$ identifies the softening of the $n=1$ mode when the sum
is continued to $\omega_0^2(\bar x)=-|U''(\bar x)|/m$ \footnote{Only $n=1$ is physically significant, marking the bifurcation of the bounce orbit from the barrier-top solution; the higher poles correspond to its $n$-fold repetitions, which are exponentially subdominant, and their presence signals that the Gaussian treatment of the unstable direction is controlled only for $T>T_0$.}. In the RCSJ model this instability is realized at the top of the barrier that the phase must overcome to escape from a metastable minimum of the tilted washboard potential, driven by the external bias current and suppressed by dissipation \cite{devoret1985, devoret1988, caldeira1981, caldeira1983, Korshunov1987}.  At the relevant bias the crossover temperature $T_0$, of the order of a few tens of mK for the parameters of Sec.~\ref{sec:rcsj}, is well above the typical operating temperature, which allows the observation of macroscopic quantum tunneling.

The second effect, closer in spirit to the Schwinger mechanism, arises when the background is itself time dependent: Eq.~\eqref{eq:fluctuations} is then a damped Hill equation, whose parametric resonances describe the amplification of quantum fluctuations by the oscillating background, with $\gamma$ setting a finite threshold \cite{saha2020}. Here the quantum corrections, besides modulating the stiffness through $U''_{\rm eff}(\bar x)$, make the friction coefficient itself background dependent, $\gamma\to\gamma+m'_{\rm eff}(\bar x)\dot{\bar x}$, a parametric modulation of the damping that has no classical counterpart for a constant bare mass and that shifts the instability threshold at the same order in $\hbar$. Related vacuum-instability phenomena driven by external fields have been investigated in spin and magnon systems \cite{Hongo2021, Yuan2023, Adorno2024}. Extending the present dissipative effective action from a single collective coordinate to the full relative-phase field $\Phi(x,t)$ of Appendix \ref{app:B}, which supports a genuine continuum of modes, would turn parametric amplification into a true pair-creation problem, with a Bogoliubov spectrum of created quanta.

We also note that $U_1(\bar x)$ is itself a Casimir-type quantity: it is the energy of the quantum fluctuations around the background, and $-U_1'(\bar x)$ the force it exerts, with the role of the geometric separation being played here by $\bar x$. Indeed, the argument of the logarithm in Eq.~\eqref{eq:ueff_matsubara} is $-\mathcal M(i\nu_n)/m$, the fluctuation operator continued to the imaginary axis. It thus plays the same role as the dielectric function $\varepsilon(i\nu_n)$ in the Lifshitz formula for absorbing media \cite{Lifshitz1956}, of which Eq.~\eqref{eq:ueff_matsubara} is the single-mode analog. The elongated bosonic junction offers an explicit example: the phonon bath consists of modes $k_n=\pi n/L$ in a box of length $L$, whose zero-point energy $\mathcal E_0=-\pi\hbar v_s/24L$ is the usual one-dimensional Casimir energy. Being independent of the background field, it does not enter the equation of motion; the geometry nonetheless affects the dynamics, because $\gamma \propto L$ and thus $U_1(\bar x)$ inherits a dependence on the system size through the coupling to the bath. Finite-size corrections to $\sigma(\bar x)$ beyond our continuum-limit treatment would also constitute a direct signature of the geometric dependence of the vacuum energy. Analogous zero-point energies have been studied in spin systems, both as finite-size corrections to the ground-state energy of antiferromagnets \cite{Neuberger1989, Hasenfratz1993} and, more recently, as magnonic Casimir effects \cite{Nakata2023, Nakata2024}, where dissipation has been shown to play a significant role. Here too, extending the present framework to the full relative-phase field would allow the geometric dependence of the vacuum energy to be addressed directly.

Several other directions remain open for future investigation, particularly in connection with current experiments. In both examples considered in this paper, we have restricted our analysis to the Josephson regime; extending the study to fully nonlinear dynamics would be especially interesting, as corrections to the effective mass would then play a central role. On the technical side, extending the derivative expansion beyond the weak-damping and low-temperature regime represents another important avenue for development. In addition, it would be natural to investigate the stochastic dynamics of fluctuations around the quantum-corrected deterministic evolution. A further direction concerns the structure of the dissipative kernel. Throughout we have assumed Ohmic damping, for which the retarded kernel is local in time and the classical equation of motion contains no derivatives higher than the second. The formalism is however not limited to this case. A super-Ohmic kernel, $\mathrm{Im}\,\alpha_R(\omega)=\gamma\omega+\gamma_3\omega^3$, adds a radiation-reaction term, turning Eq.~\eqref{eq:eom_class} into $m\ddot x+\gamma\dot x-\gamma_3\dddot x+U'(x)=\xi(t)$. This may be obtained, for instance, by retaining the next term in the expansion of the system-bath couplings of Appendix~\ref{app:B}, $c_n = JN(\ell k_n+\ell_3k_n^3+\cdots)$, and describes the emission of sound into the condensate, its magnitude being set by the extent of the tunneling region relative to the phonon wavelength. The one-loop construction of Sec.~\ref{sec:lpa} carries over; the Matsubara denominator of Appendix~\ref{app:LPA} becomes cubic, and Eq.~\eqref{eq:sigma_LPA} generalizes from a first to a second divided difference of the digamma function over its three roots. Similar equations occur well beyond the Josephson context, for instance in the growth of a cavitation nucleus in liquid helium \cite{Burmistrov2022}, where the collective coordinate is the bubble radius, and the effective action contains both a viscous term, Ohmic in the Matsubara frequency, and a sound-emission term proportional to $|\nu_n|^3$, which is precisely of the super-Ohmic form above. The two enter with opposite signs, so that viscosity suppresses and sound emission enhances the quantum nucleation rate. This problem is naturally formulated in terms of a coordinate-dependent mass and the classical bubble dynamics is governed by a Rayleigh-Plesset equation of the same form as Eq.~\eqref{eq:motion}. Extending our results to the super-Ohmic case would then allow us to study quantum corrections to such dynamics.

\section*{Acknowledgments}

This work is partially supported by the Project ``Frontiere Quantistiche'' (Dipartimenti di Eccellenza) of the Italian Ministry of University and Research, by ``Iniziativa Specifica Quantum'' of INFN, by the European Union-Next Generation EU within the European Quantum Flagship Project ``PASQuanS 2'', and the National Center for HPC, Big Data and Quantum Computing (Project No.~CN00000013, CN1 Spoke 10: ``Quantum Computing'').

\appendix

\section{Calculations in the local potential approximation}\label{app:LPA}

In this appendix, we are going to compute
\begin{align}\label{eq:compute_sigma}
    \sigma(\bar x) &= \int\frac{d\omega}{2\pi}\frac{\alpha_K(\omega)}{|\mathcal M(\omega)|^2}\nonumber\\
    &= \int\frac{d\omega}{2\pi}\,\hbar\coth\!\left(\frac{\hbar\omega}{2k_BT}\right)\text{Im}\,\chi(\omega)
\end{align}
in the local potential approximation, where
\begin{equation}
    \chi(\omega)=-\mathcal M^{-1}(\omega)=\frac{1}{m(\omega_0^2(\bar x)-\omega^2)-i\gamma\omega}.
\end{equation}
Using $\chi(-\omega)=\chi^*(\omega)$ and the oddness of the hyperbolic cotangent, Eq.~\eqref{eq:compute_sigma} can be rewritten as
\begin{equation}\label{eq:folded}
\sigma(\bar x)=-i\int\frac{d\omega}{2\pi}\,\hbar\coth\!\left(\frac{\beta\hbar\omega}{2}\right)\chi(\omega).
\end{equation}
By causality, $\chi(\omega)$ is analytic in the upper half plane; its poles sit at $\omega=\pm\omega_\gamma(\bar x)-i\gamma/2m$. Moreover, $\chi\sim-1/m\omega^2$ at large $|\omega|$, so the contour may be closed upwards and the integral computed by the residue theorem. The only enclosed singularities are the poles of the hyperbolic cotangent at $\omega=i\nu_n$, where $\nu_n=2\pi n/\beta\hbar$ ($n=0,1,2,\dots$) are bosonic Matsubara frequencies, at each of which $\hbar\coth(\beta\hbar\omega/2)$ has residue $2/\beta$. The pole at $n=0$ lies on the
contour and contributes one half of its residue. Collecting the residues,
\begin{align}
\sigma(\bar x)&=\frac{1}{\beta}\chi(0) +\frac{2}{\beta}\sum_{n\ge 1}\chi(i\nu_n)\nonumber\\
&=\frac{k_BT}{U''(\bar x)} +\frac{2}{m\beta}\sum_{n\ge 1} \frac{1}{\nu_n^2+\frac{\gamma}{m}\nu_n+\omega_0^2(\bar x)}\nonumber\\
&=\frac{k_BT}{U''(\bar x)} +\frac{2}{m\beta}\sum_{n\ge 1} \frac{1}{[\nu_n+\lambda_+(\bar x)][\nu_n+\lambda_-(\bar x)]},
\end{align}
where $\lambda_\pm(\bar x) = \frac{\gamma}{2m}\pm i\omega_\gamma(\bar x)$. The summation on the right-hand side can be rewritten as
\begin{align}
    &\sum_{n\ge 1} \frac{1}{(\nu_n+\lambda_+)(\nu_n+\lambda_-)}\nonumber\\
    &= \frac{1}{\lambda_- -\lambda_+}\sum_{n\ge 1}\left[\frac{1}{\nu_n+\lambda_+}-\frac{1}{\nu_n+\lambda_-}\right]\nonumber\\
    &= \frac{\beta\hbar/2\pi}{\lambda_- -\lambda_+}\sum_{n\ge 1}\left[\frac{1}{n+a_+}-\frac{1}{n+a_-}\right]\nonumber\\
    &= \frac{\beta\hbar/2\pi}{\lambda_- -\lambda_+}\sum_{n\ge 1}\left[\left(\frac{1}{n+a_+}-\frac{1}{n}\right)-\left(\frac{1}{n+a_-}-\frac{1}{n}\right)\right],
\end{align}
where $a_\pm = \beta\hbar \lambda_\pm/2\pi$. The round brackets yield the series representation of the digamma function, $\sum_{n\ge 1}(\frac{1}{n+a}-\frac{1}{n}) = -\gamma_E-\psi(1+a)$, where $\gamma_E$ is the Euler-Mascheroni constant, therefore
\begin{align}
    &\sum_{n\ge 1} \frac{1}{(\nu_n+\lambda_+)(\nu_n+\lambda_-)}\nonumber\\
    &=\frac{\beta\hbar}{2\pi}\frac{\psi(1+\frac{\beta\hbar\lambda_-}{2\pi})-\psi(1+\frac{\beta\hbar\lambda_+}{2\pi})}{\lambda_- - \lambda_+}
\end{align}
and
\begin{equation}\label{eq:sigma_LPA_app}
    \sigma(\bar x) = \frac{k_BT}{U''(\bar x)}+\frac{\hbar}{\pi m}\frac{\psi(1+\frac{\beta\hbar\lambda_-}{2\pi})-\psi(1+\frac{\beta\hbar\lambda_+}{2\pi})}{\lambda_- - \lambda_+},
\end{equation}
that is Eq.~\eqref{eq:sigma_LPA}.

In the overdamped regime ${\gamma > \gamma_c=2\sqrt{mU''(\bar x)}}$, where $\omega_\gamma$ is imaginary, by defining $\omega_\gamma(\bar x)= i\kappa_\gamma(\bar x)$, with $\kappa_\gamma(\bar x)=\sqrt{(\gamma/2m)^2-U''(\bar x)/m}$, we get $\lambda_\pm = \gamma/2m\mp \kappa_\gamma$, and the divided difference in Eq.~\eqref{eq:sigma_LPA_app} is real as it stands.

At critical damping $\gamma = \gamma_c$, where $\omega_\gamma=0$, the divided difference in Eq.~\eqref{eq:sigma_LPA_app} tends to the finite derivative $(\beta\hbar/2\pi)\psi'(1+\beta\hbar\gamma_c/4\pi m)$, hence
\begin{equation}
    \sigma(\bar x)|_{\gamma=\gamma_c} = \frac{k_BT}{U''(\bar x)} + \frac{\beta\hbar^2}{2\pi^2 m}\psi'\!\left(1+\frac{\beta\hbar\gamma_c}{4\pi m}\right).
\end{equation}

In the underdamped regime $\gamma < \gamma_c$, where $\omega_\gamma$ is real, $\lambda^*_- = \lambda_+$, and since $\psi(1+z^*) = \psi^*(1+z)$ and $\psi(1+z)=\psi(z)+1/z$, we obtain
\begin{equation}
    \sigma(\bar x) = \frac{\hbar}{\pi m \omega_\gamma(\bar x)}\text{Im}\,\psi\!\left(\frac{\beta\hbar\lambda_+(\bar x)}{2\pi}\right)-\frac{k_BT}{U''(\bar x)}.
\end{equation}

Let us analyze the low-temperature, high-temperature, and weak-damping limits of $\sigma(\bar x)$.

\emph{i. Low-temperature limit.} Since $\psi(1+z) = \ln z + \frac{1}{2z} -\frac{1}{12z^2} + \mathscr O(z^{-4})$ at large $z$, as $T \to 0$ one has $\sigma(\bar x) \simeq \frac{\hbar}{\pi m} \frac{\ln (\beta\hbar\lambda_-) - \ln(\beta\hbar\lambda_+)}{\lambda_- - \lambda_+}+\mathscr O(T^2)$, where $\mathscr O(T^2)=\frac{\pi\gamma(k_B T)^2}{3m^2\hbar\omega_0^4(\bar x)}+\cdots$. In the underdamped regime, $\lambda_-=\lambda_+^*$ and $\text{Im}\,\ln(\beta\hbar\lambda_+) = \arg \lambda_+=\tan^{-1}(2m\omega_\gamma/\gamma)$, giving
\begin{equation}\label{eq:sigma1}
    \sigma(\bar x) \simeq \frac{\hbar}{\pi m \omega_\gamma(\bar x)}\tan^{-1}\!\left[\frac{2m\omega_\gamma(\bar x)}{\gamma}\right]+\mathscr O(T^2).
\end{equation}
In the overdamped regime,
\begin{equation}
    \sigma(\bar x) \simeq \frac{\hbar}{\pi m \kappa_\gamma(\bar x)}\tanh^{-1}\!\left[\frac{2m\kappa_\gamma(\bar x)}{\gamma}\right]+\mathscr O(T^2).
\end{equation}
The two expressions match at critical damping $\gamma=\gamma_c$, where $\sigma(\bar x) = 2\hbar/\pi \gamma_c + \mathscr O(T^2)$.

\emph{ii. High-temperature limit.} Since ${\psi(1+z)} = -\gamma_E+\zeta(2)z-\zeta(3)z^2+\mathscr O(z^3)$ at small $z$, where $\zeta(n)$ is the Riemann zeta function, as $T \to \infty$ we have
\begin{equation}\label{eq:sigma_highT}
    \sigma(\bar x) \simeq \frac{k_BT}{U''(\bar x)} + \mathscr O(T^{-1}),
\end{equation}
where $\mathscr O(T^{-1})=\frac{\hbar^2}{12 m k_B T}-\frac{\zeta(3)}{4\pi^3}\frac{\hbar^3 \gamma}{m^2(k_B T)^2}+\cdots$. The leading term gives the classical equipartition of energy, $\frac{1}{2}U''(\bar x)\sigma(\bar x) = \frac{1}{2} k_B T$.

\emph{iii. Weak-damping limit.} As $\gamma \to 0$, $\lambda_+ \to i\omega_0$, and the identity $\text{Im}\,\psi(iy)=\frac{1}{2y}+\frac{\pi}{2}\coth(\pi y)$ yields
\begin{equation}
    \sigma(\bar x) \simeq \frac{\hbar}{2 m \omega_0(\bar x)}\coth\!\left[\frac{\beta\hbar\omega_0(\bar x)}{2}\right].
\end{equation}
This is consistent with Eqs.~\eqref{eq:sigma1} and \eqref{eq:sigma_highT} in their common limits.

Equivalently, one may consider directly the limits of the effective potential \eqref{eq:ueff}.

\emph{i. Low-temperature limit.} Since $\ln \Gamma(1+z) = z \ln z - z + \frac{1}{2}\ln(2\pi z) + \frac{1}{12 z} + \mathscr O(z^{-3})$ at large $z$, as $T \to 0$ one has $U_1(\bar x) = -\frac{\hbar}{2\pi}[\lambda_+ \ln (\lambda_+/\omega_c) + \lambda_- \ln (\lambda_-/\omega_c)]+\mathscr O(T^2)$ up to $\bar x$-independent terms, where $\mathscr O(T^2)=-\frac{\pi\gamma(k_B T)^2}{6m\hbar\omega_0^2(\bar x)}+\cdots$ and $\omega_c$ is a reference frequency scale needed on dimensional grounds. In the underdamped regime, $U_1(\bar x) = -\frac{\hbar}{\pi}\text{Re}[\lambda_+ \ln (\lambda_+/\omega_c)]+\mathscr O(T^2)$, and since $\lambda_+=\omega_0 e^{i \arg \lambda_+}$ with $\arg \lambda_+=\tan^{-1}(2m\omega_\gamma/\gamma)$, we obtain
\begin{align}
    U_\text{eff}(\bar x) &= U(\bar x)\nonumber\\
    &+ \frac{\hbar}{\pi}\left[\omega_\gamma(\bar x)\tan^{-1}\!\left(\frac{2m\omega_\gamma(\bar x)}{\gamma}\right)-\frac{\gamma}{2m}\ln\frac{\omega_0(\bar x)}{\omega_c}\right]\nonumber\\
    &+\mathscr O(T^2).
\end{align}
In the overdamped regime,
\begin{align}
    U_\text{eff}(\bar x) &= U(\bar x)\nonumber\\
    &- \frac{\hbar}{\pi}\left[\kappa_\gamma(\bar x)\tanh^{-1}\!\left(\frac{2m\kappa_\gamma(\bar x)}{\gamma}\right)+\frac{\gamma}{2m}\ln\frac{\omega_0(\bar x)}{\omega_c}\right]\nonumber\\
    &+\mathscr O(T^2).
\end{align}

\emph{ii. High-temperature limit.} Since $\ln \Gamma(1+z) = -\gamma_E z + \frac{\zeta(2)}{2}z^2-\frac{\zeta(3)}{3}z^3 + \mathscr O(z^4)$ for small $z$, 
\begin{equation}
    U_\text{eff}(\bar x) = U(\bar x)+k_BT \ln \frac{\hbar\omega_0(\bar x)}{k_BT} + \mathscr O(T^{-1}),
\end{equation}
where $\mathscr O(T^{-1}) = \frac{\hbar^2\omega_0^2(\bar x)}{24 k_B T} - \frac{\zeta(3)}{8\pi^3}\frac{\hbar^3\gamma \omega_0^2(\bar x)}{m(k_B T)^2}+\cdots$.

\emph{iii. Weak-damping limit.} As $\gamma \to 0$, $\lambda_\pm \to \pm i\omega_0$, and the identity $\Gamma(1+iy)\Gamma(1-iy)=\pi y/\sinh(\pi y)$ yields \begin{equation}
    U_{\text{eff}}(\bar x) \simeq U(\bar x) + \frac{\hbar\omega_0(\bar x)}{2} + \frac{1}{\beta}\ln\left(1-e^{-\beta\hbar\omega_0(\bar x)}\right).
\end{equation}

\section{Keldysh action for the elongated bosonic junction}\label{app:B}

In this appendix, we derive the Keldysh action for an elongated bosonic junction. This also serves as a general illustration of how the characteristic structure of the Keldysh action \eqref{eq:Kac}, the fluctuation-dissipation relation \eqref{eq:FT}, and the Markovian retarded kernel \eqref{eq:ohm} emerge from a microscopic description upon integrating out environmental degrees of freedom. 

We start from the Lagrangian density for two weakly coupled one-dimensional Bose gases of length $L$,
\begin{align}
    \mathscr L = &\sum_{j=1,2} \left(i\hbar \Psi_j^* \dot\Psi_j-\frac{\hbar^2}{2m}|\partial_x\Psi_j|^2-\frac{g}{2}|\Psi_j|^4\right)\nonumber\\
    &+ Jw(x)\left(\Psi_1^*\Psi_2+\Psi_2^*\Psi_1\right),
\end{align}
where $g = UL>0$ is the contact interaction strength, while $Jw(x)$, with $J$ the tunneling energy, describes the coupling between the two components. We shall consider a head-to-tail configuration, in which the dimensionless profile $w(x)$ is localized near the edge at $x=0$, with normalization $\int_0^L \frac{dx}{L}\,w(x)=1$. Expressing the bosonic fields in density-phase form, $\Psi_j(x,t)=\sqrt{N_j(x,t)/L}\,e^{i\Phi_j(x,t)}$, we obtain, up to total derivatives,
\begin{align}
    \mathscr L = \frac{1}{L}\sum_{j=1,2} \biggl[&-\hbar N_j \dot\Phi_j - \frac{\hbar^2}{2m}\left(\frac{(\partial_x N_j)^2}{4 N_j}+N_j(\partial_x\Phi_j)^2\right) \nonumber\\
    &-\frac{U}{2}N_j^2\biggr] + \frac{2J w(x)}{L}\sqrt{N_1 N_2}\cos(\Phi_2-\Phi_1).
\end{align}
We now introduce the space-time dependent relative phase $\Phi(x,t) = \Phi_2(x,t) - \Phi_1(x,t)$ and fractional population imbalance $Z(x,t) = [N_1(x,t) - N_2(x,t)]/N$, where $N$ is the (constant) total number of particles. Assuming negligible gradients of the total phase, $\partial_x(\Phi_1+\Phi_2)\simeq 0$, meaning that the junction carries no significant net mass current, we then obtain
\begin{align}\label{eq:lag_app}
    \mathscr L \simeq &\,\frac{N\hbar Z}{2L}\dot\Phi -\frac{\hbar^2N}{8mL}\left[\frac{(\partial_xZ)^2}{1-Z^2}+(\partial_x\Phi)^2\right]\nonumber\\
    &-\frac{UN^2}{4L}Z^2 + \frac{JN w(x)}{L}\sqrt{1-Z^2}\cos\Phi.
\end{align}
We next expand the fields $\Phi(x,t)$ and $Z(x,t)$ in eigenmodes of the Laplacian, namely functions $\chi_n(x)$ satisfying $-\partial_x^2\chi_n(x)=k_n^2\chi_n(x)$. On the interval $x \in [0,L]$ with vanishing boundary conditions, the normalized eigenfunctions are $\chi_n(x)=\sqrt 2 \sin(k_nx)$, with $k_n=\pi n/L$ for $n=1,\dots,\infty$, and satisfy $\int dx\,\chi_n(x)\chi_m(x) = L\delta_{nm}$. In addition to these bulk modes, which correspond to longitudinal phonon excitations, we have the boundary fields $\phi(t)=\Phi(0,t)$ and $z(t)= Z(0,t)$, which represent the Josephson degree of freedom. We therefore write
\begin{align}
    \Phi(x,t) &= \phi(t)+\sum_{n=1}^\infty q_n(t)\chi_n(x),\\
    Z(x,t) &= z(t)+\sum_{n=1}^\infty \frac{2p_n(t)}{N\hbar}\chi_n(x).
\end{align}
Substituting these expressions into Eq.~\eqref{eq:lag_app}, and further assuming that the system remains nearly homogeneous in space, with imbalance fluctuations small compared with the background imbalance, so that $(\partial_xZ)^2 \simeq 0$ and $\sqrt{1-Z^2}\cos\Phi \simeq \sqrt{1-z^2}\cos\Phi$, we arrive at the action
\begin{align}\label{eq:ac_CL}
    S \simeq &\int dt\biggl[\frac{N\hbar z}{2}\dot\phi - \frac{UN^2}{4}z^2 \nonumber\\
    &+ JN\sqrt{1-z^2}\int_0^L\frac{dx}{L}\,w(x)\cos\left(\phi+\sum_{n=1}^\infty \chi_n(x) q_n\right)\nonumber\\
    &+\sum_{n=1}^\infty\left( p_n \dot q_n - \frac{p_n^2}{2M}-\frac{M\omega_n^2}{2} q_n^2\right)\biggr],
\end{align}
where we defined $M = \hbar^2/2U$ and $\omega_n = \sqrt{UN/2m}\,k_n = v_sk_n$, with $v_s$ the Bogoliubov speed of sound.

The action \eqref{eq:ac_CL} describes the intrinsic coupling between the Josephson mode and the phonon excitations supported by the elongated Bose gases. Expanding the tunneling term to first order in the phase fluctuations, and further linearizing the coupling with $q_n$ under the assumption of small $\phi(t)$ and $z(t)$, we get $JN\sqrt{1-z^2}\int_0^L \frac{dx}{L}\,w(x)\cos(\phi + \sum_{n=1}^\infty \chi_n(x) q_n) \simeq JN\sqrt{1-z^2}\cos \phi -\sum_{n=1}^\infty c_n \phi q_n$, where 
\begin{equation}\label{eq:c_n}
    c_n = JN\int_0^L \frac{dx}{L}\,w(x)\chi_n(x).
\end{equation}
In this way, we obtain a linear system-bath coupling of the Caldeira-Leggett type \cite{caldeira1981, caldeira1983}, where the bath is represented by a collection of uncoupled oscillators:
\begin{equation}\label{eq:dissa}
    S \simeq S_0[\phi,z] + \sum_{n=1}^\infty S_n[\phi, q_n],
\end{equation}
where
\begin{align}
    S_0 &= \int dt \left[\frac{N\hbar z}{2}\dot\phi-H(\phi,z)\right],\\
    H &= \frac{UN^2}{4}z^2-JN\sqrt{1-z^2}\cos\phi,
\end{align}
and
\begin{align}\label{eq:sn}
    S_n &= \int dt\left[\frac{M}{2}\dot q_n^2-\frac{M\omega_n^2}{2}q_n^2 - c_n \phi q_n\right].
\end{align}

The objective is to now integrate out the bath degrees of freedom to obtain the Keldysh action for the Josephson mode. Introducing forward and backward trajectories, the Keldysh representation of the action \eqref{eq:dissa} is
\begin{equation}
    S_K = S_\text{cons}[\phi_\pm, z_\pm] + \sum_{n=1}^\infty S_n[\phi_\pm, q_{n,\pm}],
\end{equation}
where $S_\text{cons} = S_0[\phi_+, z_+] - S_0[\phi_-, z_-]$ and $S_n = S_n[\phi_+, q_{n,+}]-S_n[\phi_-, q_{n,-}]$. The Keldysh effective action for the Josephson mode is then defined as
\begin{align}\label{eq:pib}
    &e^{iS_K^{(\text{eff})}[\phi_\pm, z_\pm]} = e^{iS_\text{cons}[\phi_\pm, z_\pm]}\prod_{n=1}^\infty\int dq^i_{n,+} dq^i_{n,-} dq_n^f\nonumber\\
    &\times \rho_B(q^i_{n,+}, q^i_{n,-}) \int_{q^i_{n,+}}^{q^f_n} \mathcal Dq_{n,+} \int_{q^i_{n,-}}^{q^f_n} \mathcal Dq_{n,-}\,e^{iS_n[\phi_\pm, q_{n,\pm}]},
\end{align}
where $\rho_B(q^i_{n,+}, q^i_{n,-}) = \langle q^i_{n,+} | \hat\rho_B | q^i_{n,-}\rangle$ is the density matrix element weighting the initial conditions for the bath. Assuming the bath to be in thermal equilibrium, this is given by \cite{feynman_book}
\begin{align}
    &\rho_B(q^i_{n,+}, q^i_{n,-}) \propto \exp \biggl[-\frac{M\omega_n}{2\hbar \sinh(\beta\hbar \omega_n)}\nonumber\\
    &\times \left[\left((q^i_{n,+})^2+(q^i_{n,-})^2\right) \cosh(\beta\hbar\omega_n)-2q^i_{n,+}q^i_{n,-}\right]\biggr].
\end{align}
Performing the Keldysh rotation to classical and response fields, we obtain
\begin{align}
    S_\text{cons} &= \int dt \biggl[\left(\frac{N\hbar}{2}\dot\phi-\frac{UN^2}{2}z-\frac{JNz}{\sqrt{1-z^2}}\cos\phi\right)\!z_q\nonumber\\
    &- \left(\frac{N\hbar}{2}\dot z + JN \sqrt{1-z^2}\sin\phi\right) \!\phi_q\biggr] + \mathscr O(\phi_q^3, z_q^3),\\
    S_n &= \int dt\left[-(M \ddot q_n + M\omega_n^2q_n + c_n\phi)q_{n,q} - c_n q_n \phi_q\right],
\end{align}
and
\begin{align}
    \rho_B(q^i_n, q^i_{n,q}) &\propto \exp\biggl[-\frac{M\omega_n}{\hbar}\tanh\left(\frac{\beta\hbar\omega_n}{2}\right)(q^i_n)^2\nonumber\\
    &-\frac{M\omega_n}{4\hbar}\coth\left(\frac{\beta\hbar\omega_n}{2}\right)(q^i_{n,q})^2\biggr].
\end{align}
Eq.~\eqref{eq:pib} then reads
\begin{align}
    &e^{iS_K^{(\text{eff})}} = e^{iS_\text{cons}}\prod_{n=1}^\infty\int dq_n^i\,dq_{n,q}^i\, dq_n^f\nonumber\\
    &\times \rho_B(q^i_n, q^i_{n,q}) \int_{q_n^i}^{q_n^f} \mathcal Dq_n \int_{q_{n,q}^i}^0 \mathcal Dq_{n,q}\,e^{iS_n}.
\end{align}
Since $S_n$ is linear in $q_{n,q}$, the rightmost integral yields $\int \mathcal Dq_{n,q} e^{iS_n}=e^{-i\int dt\,c_n q_n\phi_q} \delta(M\ddot q_n+M\omega_n^2 q_n + c_n\phi)$. Then, integrating over $q_n$, the functional delta enforces the EoM of Eq.~\eqref{eq:sn}, whose solution is
\begin{align}
    q_n(t) = q_n^{\text{(h)}}(t) - \frac{1}{c_n}\int dt' \alpha_{n,R}(t-t')\phi(t'),
\end{align}
where $q_n^{\text{(h)}}(t) = q_n^i\cos(\omega_n t)+\frac{\dot q_n^i}{\omega_n}\sin(\omega_nt)$ is the homogeneous solution, and
\begin{equation}
    \alpha_{n,R}(t-t') = \Theta(t-t') \frac{c_n^2}{M\omega_n}\sin[\omega_n(t-t')].
\end{equation}
It follows that
\begin{align}\label{eq:sk}
    e^{iS_K^{(\text{eff})}} &= e^{iS_\text{cons}}\prod_{n=1}^\infty e^{i\int dt\,dt' \phi_q(t)\alpha_{n,R}(t-t')\phi(t')}\nonumber\\
    &\times \int dq_n^i\,\rho_B(q^i_n, 0)\,e^{-i c_n\int dt\,q_n^{(\text h)} \phi_q}.
\end{align}
In the last line, we have the expectation value $\langle e^{-i c_n\int dt\,q_n^{(\text h)} \phi_q}\rangle_{\rho_B}$, evaluated on the equilibrium density matrix of a harmonic oscillator. Since $\rho_B$ is a zero-mean Gaussian, this is equivalent to $e^{-\frac{c_n^2}{2} \int dt\,dt' \phi_q(t) \langle q_n^{(\text h)}(t) q_n^{(\text h)}(t')\rangle_{\rho_B}\phi_q(t')}$. Using $\langle(q_n^i)^2\rangle_{\rho_B} = \frac{\hbar}{2M\omega_n}\coth(\frac{\beta\hbar\omega_n}{2})$, $\langle(\dot q_n^i)^2\rangle_{\rho_B} = \omega_n^2\langle(q_n^i)^2\rangle_{\rho_B}$, and $\langle q_n^i \dot q_n^i\rangle_{\rho_B} = 0$, we obtain $\langle q_n^{(\text h)}(t) q_n^{(\text h)}(t')\rangle_{\rho_B} = \frac{\hbar}{2M\omega_n}\coth(\frac{\beta\hbar\omega_n}{2})\cos[\omega_n(t-t')]$. Substituting this back into Eq.~\eqref{eq:sk}, we finally get
\begin{align}\label{eq:skap}
    S_K = S_\text{cons} &+ \int dt\,dt'\biggl[ \phi_q(t) \alpha_R(t-t')\phi(t')\nonumber\\
    &+\frac{i}{2}\phi_q(t) \alpha_K(t-t') \phi_q(t')\biggr],
\end{align}
where
\begin{align}
    \alpha_R(t-t') &= \Theta(t-t') \sum_{n=1}^\infty \frac{c_n^2}{M\omega_n}\sin[\omega_n(t-t')],\\
    \alpha_K(t-t') &=\hbar\sum_{n=1}^\infty \frac{c_n^2}{2M\omega_n}\coth\left(\frac{\beta\hbar\omega_n}{2}\right) \cos[\omega_n(t-t')].
\end{align}
The structure of the action \eqref{eq:skap} is exactly the one given by Eqs.~\eqref{eq:Kac}-\eqref{eq:acK_diss}. Defining the spectral density 
\begin{equation}\label{eq:specdens}
    \mathcal J(\omega) = \frac{\pi}{2}\sum_{n=1}^\infty \frac{c_n^2}{M\omega_n}\delta(\omega-\omega_n),\quad \omega>0,
\end{equation}
we may write the kernels as
\begin{align}
    \alpha_R(t-t') &= \Theta(t-t') \int \frac{d\omega}{2\pi}\,4\mathcal J(\omega) \sin[\omega(t-t')],\label{alphar_def}\\
    \alpha_K(t-t') &= \hbar \int \frac{d\omega}{2\pi}\,2\mathcal J(\omega)\coth\left(\frac{\beta\hbar\omega}{2}\right)\cos[\omega(t-t')].
\end{align}
From this we find
\begin{equation}
    \alpha_K(\omega) = \hbar \coth\left(\frac{\beta\hbar\omega}{2}\right)\text{Im}\,\alpha_R(\omega),
\end{equation}
that is the fluctuation-dissipation relation \eqref{eq:FT}.

It remains only to derive the explicit expression of $\mathcal J(\omega)$ for our model. Since the tunneling profile $w(x)$ is localized near $x=0$, where $\chi_n(x) \simeq \sqrt 2 k_nx$, the couplings \eqref{eq:c_n} are approximately linear in $k_n$: $c_n \simeq JN[\sqrt{2}\int_0^L \frac{dx}{L}\,x w(x)]k_n  = JN\ell k_n$. Taking the continuum limit, the spectral density \eqref{eq:specdens} is thus given by
\begin{align}
    \mathcal J(\omega) &= \frac{\pi}{2}\sum_{n=1}^\infty \frac{(JN\ell k_n)^2}{M v_sk_n}\delta(\omega-v_sk_n)\nonumber\\
    &\to \frac{(JN\ell)^2L}{2M v_s} \int_0^\infty dk\,k\,\delta(\omega-v_sk)\nonumber\\
    &= \frac{(JN\ell)^2L}{2M v_s^3}\omega = \gamma\omega,
\end{align}
that is the characteristic form of an Ohmic bath. Substituting this into Eq.~\eqref{alphar_def} and taking the Fourier transform, we obtain indeed
\begin{equation}
    \text{Im}\,\alpha_R(\omega)=\gamma\omega.
\end{equation}
The corresponding regularized $\alpha_R(t-t')$ is, in the sense of distributions,
\begin{align}
    \alpha_R(t-t') &= \Theta(t-t') 2\gamma\int_0^\infty \frac{d\omega}{\pi}\,\omega \sin[\omega(t-t')]\nonumber\\
    &=-\gamma\,\partial_t\delta(t-t'),
\end{align}
which coincides with Eq.~\eqref{eq:ohm}. 

The rescaling to adimensional units adopted in Sec.~\ref{sec:ebj} is obtained by $2Jt/\hbar \to t$ \cite{smerzi1997}.

\section{Hamilton principle for weak damping}\label{app:C}

In this appendix, we present an alternative approach to deriving the low-temperature, weak-damping EoM \eqref{eq:motion}. We are dealing with a particle of mass $m$, moving in a potential $U(x)$ and subject to a friction force $F = -\gamma\dot x$. The classical EoM can be obtained from the generalized Hamilton principle \cite{whittaker_book}
\begin{equation}\label{eq:Hamgen}
    \int dt\left[\delta\left(\frac{m}{2}\dot x^2 - U(x)\right) - \gamma\dot x\,\delta x\right] = 0.
\end{equation}
The first term is the variation of the conservative action \eqref{eq:s0}, namely the exact differential $\delta S_0 = \int dt\,\frac{\delta S_0}{\delta x}\delta x$, with functional derivative $\frac{\delta S_0}{\delta x} = -m\ddot x-U'(x)$. In the second term, $\delta W = -\gamma\dot x\,\delta x$ is the infinitesimal virtual work done by the nonconservative friction force and, unlike $\delta S_0$, is not an exact differential. Introducing the functional $\mathscr E[x]$ such that Eq.~\eqref{eq:Hamgen} can be written as $\int dt\,\mathscr E[x] \delta x =0$, the arbitrariness of $\delta x$ implies the EoM
\begin{equation}
    \mathscr {E}[x]  = -m\ddot x-U'(x)-\gamma\dot x=0.
\end{equation}
The corresponding one-loop quantum-corrected EoM is
\begin{equation}
    \mathscr E[\bar x] + \frac{i\hbar}{2}\frac{\delta}{\delta\bar x}\,\text{tr} \ln\left(\frac{\delta\mathscr E}{\delta\bar x}\right)=0,
\end{equation}
where
\begin{equation}
    \frac{\delta\mathscr E}{\delta\bar x} = -m\partial_t^2-\gamma\partial_t-U''(\bar x)
\end{equation}
is precisely the fluctuation operator $\mathcal M$ introduced in Eq.~\eqref{eq:M}. In frequency space, this operator takes the form $\mathcal M(\omega) = m\omega^2 + i\gamma\omega-U''(\bar x)=m[(\omega+\frac{i\gamma}{2m})^2-\omega^2_\gamma(\bar x)]$, with $\omega_\gamma(\bar x)$ given by Eq.~\eqref{eq:omegaeff}. In the weak damping limit, the small imaginary shift of the zeros can be neglected, which yields
\begin{equation}
    \mathscr E[\bar x] + \frac{i\hbar}{2}\frac{\delta}{\delta\bar x}\int\frac{d\omega}{2\pi} \ln[\omega^2-\omega_\gamma^2(\bar x)] = 0.
\end{equation}
This reproduces the quantum-corrected EoM \eqref{eq:motion}. 

\bibliography{References}

\end{document}